\documentclass{aa}  
\usepackage{natbib}
\usepackage{graphicx}
\usepackage{gensymb}
\usepackage[toc,page]{appendix}
\usepackage{pbox}
\usepackage{makecell}
\usepackage[flushleft]{threeparttable}

\usepackage[flushleft]{threeparttable}

\usepackage{hyperref}
\usepackage{multirow}
\usepackage{multicol}
\usepackage{longtable,tabu}   
\hypersetup{colorlinks=true,breaklinks=true,citecolor=blue}
\usepackage{txfonts}

\bibpunct{(}{)}{;}{a}{}{,} 
\begin{document}

   \title{Broad line region and black hole mass of PKS 1510-089 from spectroscopic reverberation mapping}

   \author{Suvendu Rakshit 
          }

   \institute{Finnish Centre for Astronomy with ESO (FINCA), 
University of Turku, Quantum, Vesilinnantie 5, 20014, Finland\\ \email{suvenduat@gmail.com} \\
             }

   \date{-}


\abstract
   {
Reverberation results of a flat spectrum radio quasar PKS 1510-089 are presented from 8.5-years long spectroscopic monitoring carried out in 9 observing seasons between December 2008 to June 2017 at Steward Observatory. Optical spectra show strong H$\beta$, H$\gamma$, and Fe II emission lines overlaying on a blue continuum. All the continuum and emission line light curves show significant variability with a fractional root-mean-square variation of $37.30\pm0.06$\% ($f_{5100}$), $11.88\pm0.29$\% (H$\beta$) and $9.61\pm0.71$\% (H$\gamma$), however, along with thermal radiation from accretion disk non-thermal emission from jet also contribute to $f_{5100}$. Several methods of time series analysis (ICCF, DCF, von Neumann, Bartels, \textsc{javelin}, $\chi^2$) are used to measure lag between continuum and line light curves. The observed frame BLR size is found to be $61.1^{+4.0}_{-3.2}$ ($64.7^{+27.1}_{-10.6}$) light-days for H$\beta$ (H$\gamma$). Using $\sigma_{\mathrm{line}}$ of $1262\pm247$ km s$^{-1}$ measured from the rms spectrum, the black hole mass of PKS 1510-089 is estimated to be $5.71^{+0.62}_{-0.58} \times 10^{7} M_{\odot}$.
   }


\titlerunning{Broad line region and black hole mass of PKS 1510-089}
   \maketitle
%

\section{Introduction}
Active galactic nuclei (AGN) are powered by the accretion of matter on a central supermassive black hole surrounded by an accretion disk and broad line region \citep[BLR; see][]{1995PASP..107..803U}. The BLR is photo-ionized by the UV/optical photon from the accretion disk emitting broad emission lines of FWHM ($10^{3}-10^{5}$ km/s), that are detected through optical spectroscopy. The mass of the black hole is found to be strongly correlated with the host galaxy properties suggesting that the co-evolution of the black hole and host galaxy \citep{2013ARA&A..51..511K}. However, accurate measurement of black hole masses is crucial. The black hole masses can be dynamically measured in the nearby galaxy using star and gas, however, this is extremely challenging for AGN beyond the local volume. More challenging is the measurement of black hole mass in radio-loud AGNs since their optical emission is dominated by the non-thermal emission from their relativistic jets that are aligned close to the observer.

PKS 1510–089 is a well-studied flat spectrum radio quasar (FSRQ) located at a redshift $z=0.361$ \citep{1990PASP..102.1235T}. Its optical spectrum shows broad emission lines with a blue continuum \citep{1993MNRAS.263..999T}. The broad-band spectral energy distribution (SED) of FSRQs has a double hump structure due to the combined effect of the thermal emission from accretion disk peaking at UV and optical, and non-thermal emission from the jets \citep[e.g.,][]{2010ApJ...716...30A}. The low energy (radio to X-rays) is dominated by the optically thin synchrotron emission from relativistic electron from jet while the high energy peak (X-ray to $\gamma$-ray) could be due to inverse Compton process where the seed photons originating from the BLR \citep{1994ApJ...421..153S}. Similar to other radio-loud AGNs, it shows strong flux variation across the entire electromagnetic spectrum from radio to $\gamma$-ray \citep[e.g.,][]{1986ApJ...300..216M,2000ApJ...543..535T,2007A&A...464..175B,2010ApJ...721.1425A,2010ApJ...710L.126M,2013MNRAS.428.2418O,2013A&A...554A.107H,2016ApJ...822L..13K,2017A&A...606A..87B,2017A&A...601A..30C,2019ApJ...883..137P}. However, the broad line region of PKS 1510–089 remains least studied due to its strong radio emission although the size of BLR is an important model parameter in the multi-band SED fitting. Moreover, the mass of the black hole powering PKS 1510–089 remains highly uncertain \citep{2010ApJ...721.1425A}.

Reverberation mapping \citep[RM;][]{1982ApJ...255..419B,1993PASP..105..247P} is a reliable tool to estimate the size of the BLR and black hole mass through spectroscopic monitoring. It has so far provided BLR size of more than 100 objects \citep{1999ApJ...526..579W,2000ApJ...533..631K,2004ApJ...613..682P,2009ApJ...697..160B,2013ApJ...767..149B,2016ApJ...818...30S,2017ApJ...851...21G,2017ApJ...847..125P,2016ApJ...825..126D,2018ApJ...856....6D,2019ApJ...886...93R,2020ApJ...892...93C}  allowing us to establish a relation between the size of the BLR and monochromatic luminosity \citep{2013ApJ...767..149B,2019ApJ...886...42D}. Recently, RM studies of a few FSRQs have been performed using multi-year monitoring data. For example, \citet{2019ApJ...876...49Z} performed a reverberation study of the FSRQ 3C273 using 10-yr long Steward Observatory monitoring data providing a highly reliable measurement of BLR size. \citet{2019A&A...631A...4N} performed Mg II reverberation studies of FSRQ 3C 454.3 using the Steward Observatory monitoring data and \citet{zajaek2020timedelay} measured Mg II lag of HE 0413-4031 using SALT monitoring data.

As a part of the optical spectropolarimetric monitoring program of $\gamma-$ray emitting Blazar, PKS 1510–089 have been observed in Steward Observatory from 2008 with a median time sampling of $\sim$10 days \citep{2009arXiv0912.3621S}. In this paper, I analyze $\sim$8.5-years (total 9 observing seasons) long optical spectroscopic data obtained from Steward Observatory. The optical spectrum shows a blue continuum and presence of strong Balmer lines (H$\beta$ and H$\gamma$) as well as Fe II emission. Both the Balmer lines show flux variability. I perform cross-correlation analysis to estimate the size of the BLR and black hole mass. This is the first reverberation based black hole mass estimates of PKS 1510–089. In section \ref{sec:data}, I describe the data analysis and in section \ref{sec:result} I present the result of the spectral analysis. I briefly discuss the result in section \ref{sec:discussion} and concluded the work in section \ref{sec:conclusion}.

\section{Data}\label{sec:data}

\subsection{Optical data}\label{see:optical_data}
For this work, optical photometry and spectroscopic data from Steward Observatory spectropolarimetric monitoring project\footnote{\url{http://james.as.arizona.edu/~psmith/Fermi/}}, a support program for the Fermi Gamma-Ray Space Telescope, was used. Observations were carried out using spectrophotometric instrument SPOL \citep{1992ApJ...398L..57S} with a 600 mm$^{-1}$ grating, which provides a wavelength coverage of $4000-7550 \AA$ and a spectral resolution of $15-25\AA$ depending on the slit width \citep{smith2009coordinated}. Observations were performed using the 2.3 m Bok Telescope on Kitt Peak and the 1.54 m Kuiper Telescope on Mt. Bigelow in Arizona. Details of observations and data reduction are given in \citet{smith2009coordinated}. In short, differential photometry using a standard field star was preformed to calibrate photometric magnitudes. There were 363 $V$-band photometric observations carried out between December 2008 and July 2017, used in the work. Spectra were flux calibrated using the average sensitivity function which was derived from multiple observations of several spectrophotometric standard stars throughout an observing campaign. Final flux calibrations were performed rescaling the nightly spectrum to match the synthetic $V$-band photometry of that night \citep[see][]{smith2009coordinated}. Therefore, a total of 341 photometrically calibrated spectra obtained between December 2008 and June 2017 were downloaded from Steward Observatory database and used in this work.

     \begin{figure}
                  \resizebox{9cm}{7cm}{\includegraphics{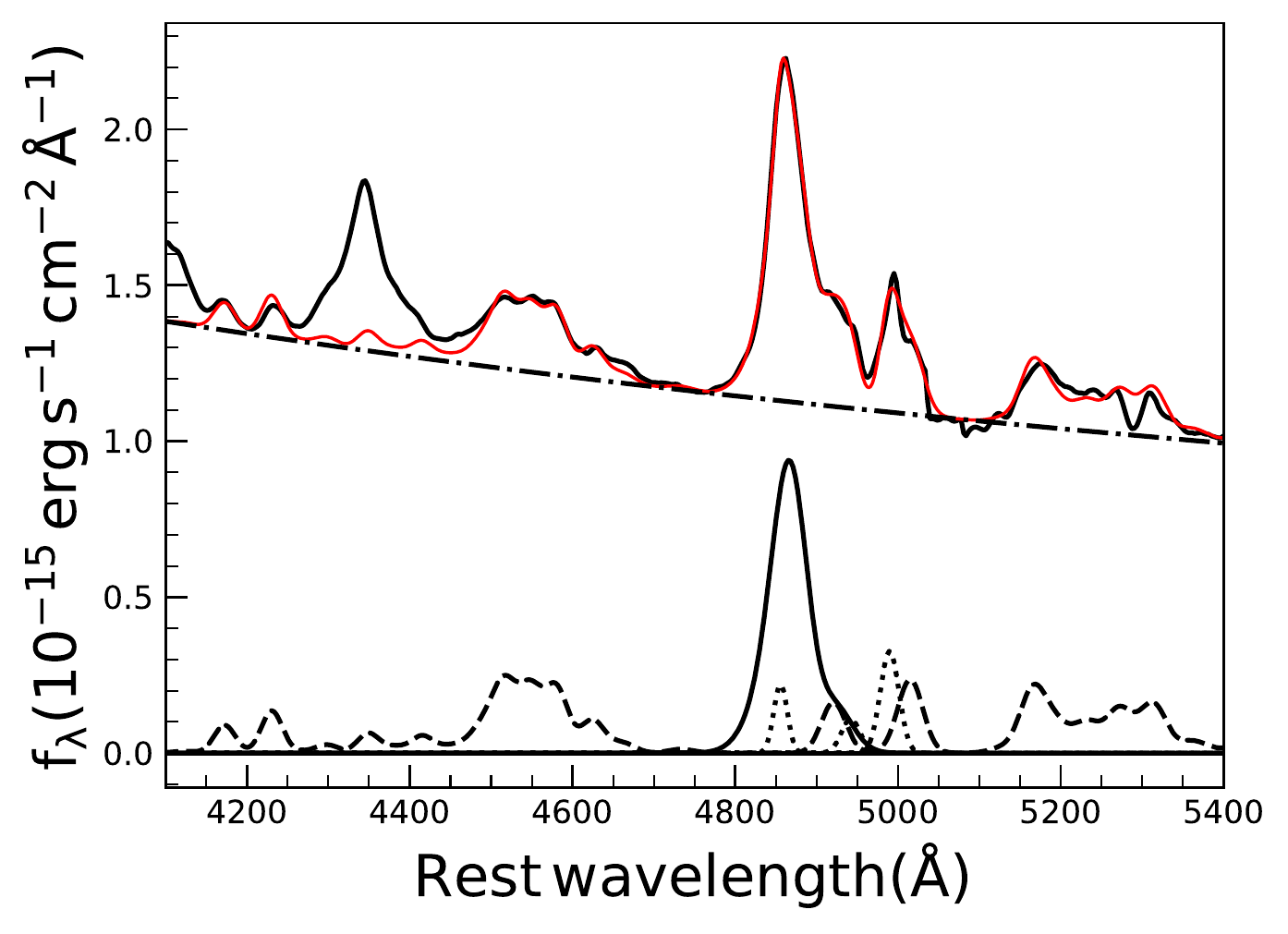}}  
                  \caption{Example of the spectral decomposition of PKS 1510-089. The rest-frame spectrum (black), best-fit model (red), decomposed AGN power-law component (dashed-dot) are shown along with the Fe II emission (dashed), broad H$\beta$ (solid) and narrow H$\beta$ and [O III] (dotted).}\label{Fig:model} 
     \end{figure}    

\subsection{Gamma-ray and Radio data}

The $\gamma$-ray data were collected from the publicly available database of Large Area Telescope (LAT) on board of the Fermi Gamma-Ray Space Telescope \citep{2009ApJ...700..597A} between December 2008 to June 2017 within the energy range of 100 MeV to 300 GeV. For data analysis, Fermi Science Tool version v10r0p5 and publicly available \textsc{fermipy} package \citep{wood2017fermipy} was used. The data sets within 15$\degree$ of the region of interest and a zenith angle cut of more than 90$\degree$ were considered to avoid background contamination. The instrument response function `P8R2\_SOURCE\_V6', the isotropic background
model `iso\_P8R2\_SOURCE\_V6\_v06.txt' and the Galactic diffuse model `gll\_iem\_v06.fit' were also used. The analysis was performed using the maximum likelihood method (`gtlike') with the criteria `(DATA\_QUAL $>$
0)\&\&(LAT\_CONFIG==1)' and the monthly binned light curve was generated.

The 15 GHz radio data observed using 40 M Telescope at the Owens Valley Radio Observatory (OVRO) was also collected. The data was obtained as a part of an observation program supporting the Fermi Gamma-ray Space Telescope having a time sampling of about twice per week \citep{2011ApJS..194...29R}.

           \begin{figure*}
           \centering
          \resizebox{12cm}{19cm}{\includegraphics{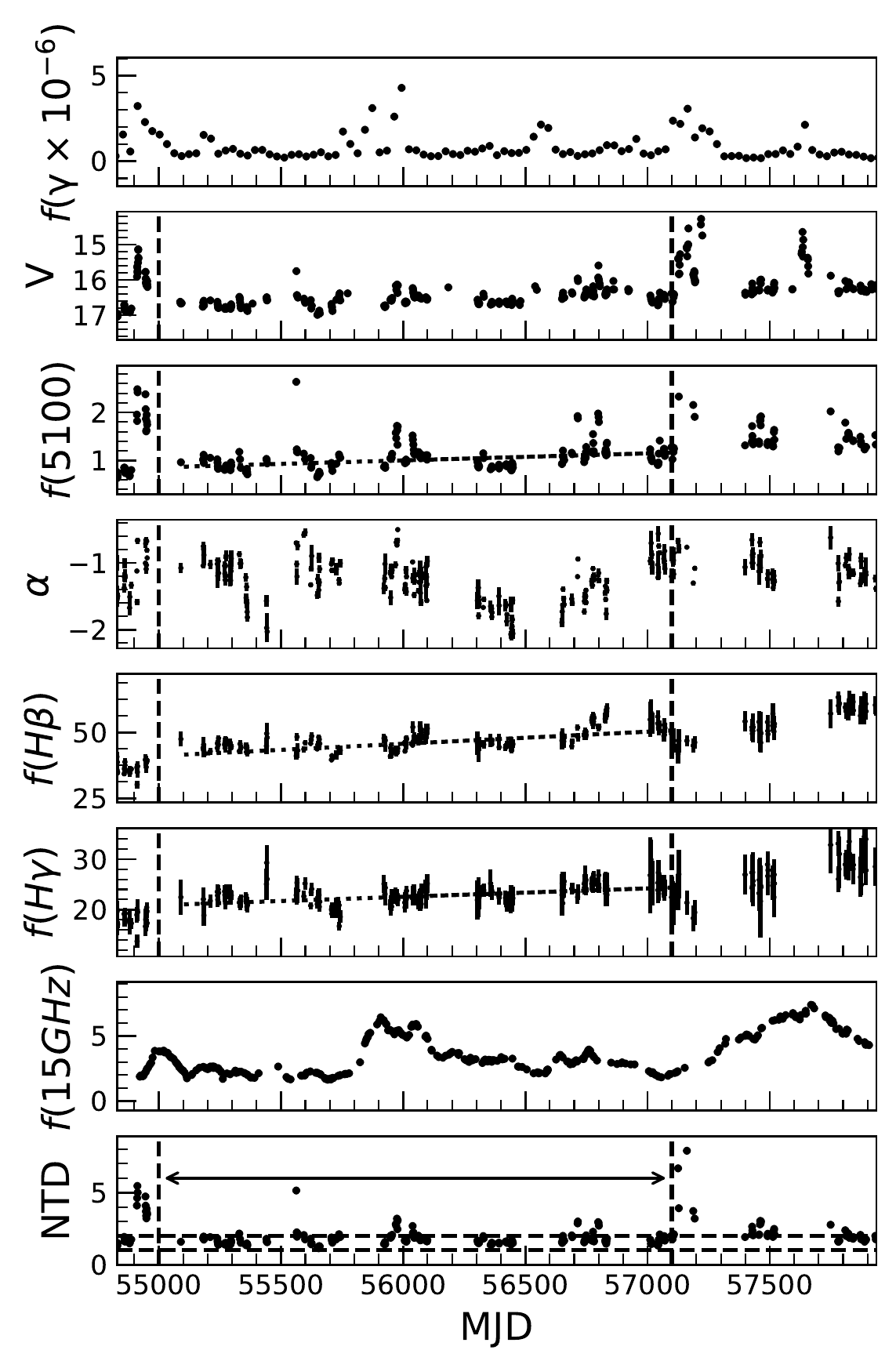}}  
          \caption{Light curves of PKS 1510-089. From top to bottom, the variation of $\gamma$-ray, $V$-band, 5100 \AA\, continuum, spectral index, H$\beta$, H$\gamma$, radio, and NTD (see text) with time are shown. The unit of $\gamma$-ray flux is photons s$^{-1}$ cm$^{-2}$, $f_{5100}$ is $10^{-15}$ erg s$^{-1}$ cm$^{-2}$ \AA$^{-1}$, emission line flux is $10^{-15}$ erg s$^{-1}$ cm$^{-2}$ and radio flux is Jy. The section between MJD = 55000$-$57100 (window `A') of the spectroscopic light curves used in the time series analysis is represented by the vertical lines. The dotted-line in the spectroscopic light curve is a linear fit to the data for detrending. Two horizontal lines at NTD=1 and 2 are shown in the lower panel.}\label{Fig:lc} 
        \end{figure*}

\section{Result and analysis}\label{sec:result}
\subsection{Optical spectral decomposition}\label{sec:decomposition}

Multi-component spectral analysis is performed to obtain spectral information from each nightly spectrum. First, each spectrum is corrected for Galactic extinction using E(B-V)=0.09 \citep{2011ApJ...737..103S} and the Milky Way extinction law with $R_V = 3.1$ from \citet{1989ApJ...345..245C}. Then spectra are brought to the rest-frame using $z=0.361$. Finally, a multi-component spectral analysis is performed as done in our previous works \citep[e.g.,][]{2018ApJ...865....5R,2019ApJ...886...93R}. 

In the spectral analysis method, the continuum is first modeled with a single power-law in the form of $f_{\lambda}= \beta \lambda^{\alpha}$. In addition, to model Fe II emission, an Fe II template from \citet{2010ApJS..189...15K} is used as it provides accurate fitting of blended Fe II emission lines \citep[e.g.,][]{2017ApJ...839...93P}. During this step, all the broad and narrow emission lines are masked out. Using IDL fitting package MPFIT\footnote{\url{http://purl.com/net/mpfit}} \citep{2009ASPC..411..251M}, a nonlinear Levenberg–Marquardt least-squares minimization is performed to find the best fit continuum model. Then the best-fit continuum model is subtracted from each spectrum and used the residual spectrum to model emission lines.

  \begin{table*}
  \caption{Spectroscopic data. Columns are (1) Modified Julian Date (2) monochromatic continuum flux at 5100\AA\ in the units of $10^{-15}$ erg s$^{-1}$ cm$^{-2}$ \AA$^{-1}$, (3) and (4) are the H$\beta$ and H$\gamma$ line fluxes, respectively, in the units of $10^{-15}$ erg s$^{-1}$ cm$^{-2}$. The detrended light curves are marked by * symbol (columns 5-7). The table is available in its entirety in a machine-readable form in the online journal. A portion is shown here for guidance regarding its form and content.}
  \begin{center}
  \begin{tabular}{ccccccc} \\ \hline \hline
  MJD & $f_{5100}$ & $f(\mathrm{H\beta})$ & $f(\mathrm{H\gamma})$ & $f_{5100}^{*}$ & $f(\mathrm{H\beta})^{*}$ & $f(\mathrm{H\gamma})^{*}$ \\ 
  (1) & (2)        & (3)                  & (4)                   & (5)         & (6)                   & (7) \\    \hline
  54829 & 0.764 $\pm$ 0.010 & 40.785 $\pm$ 1.821 & 20.094 $\pm$ 1.636 & 1.123 $\pm$ 0.010 & 48.302 $\pm$ 1.821 & 20.094 $\pm$ 1.636 \\
  54830 & 0.636 $\pm$ 0.007 & 35.713 $\pm$ 2.058 & 19.649 $\pm$ 1.761 & 0.995 $\pm$ 0.007 & 43.226 $\pm$ 2.058 & 19.649 $\pm$ 1.761 \\
  54831 & 0.647 $\pm$ 0.010 & 38.696 $\pm$ 2.291 & 17.356 $\pm$ 2.461 & 1.006 $\pm$ 0.010 & 46.204 $\pm$ 2.291 & 17.356 $\pm$ 2.461  \\ \hline \hline
  \end{tabular}
  \label{Table:data}
  \end{center}
  \end{table*}

The H$\beta$ emission line complex is modeled in the wavelength range of $4740\AA - 5050 \AA$ where a sixth-order Gauss–Hermite (GH) series is used to model H$\beta$ broad component and a Gaussian to model narrow H$\beta$ component with an upper limit in the full width at half maximum (FWHM) of 1200 km s$^{-1}$. The [O III] $\lambda\lambda$4959, 5007 doublets are modeled using two Gaussian functions where an upper limit of the FWHM of the core component is set to 1200 km s$^{-1}$. During the fit, the flux ratio of [O III] $\lambda$4959 and [O III] $\lambda$5007 is fixed to its theoretical value. The spectral decomposition is applied to each nightly spectrum, an example of such decomposition is shown in Figure \ref{Fig:model}. 

To minimize any systematic uncertainty due to the decompositions of broad and narrow H$\beta$ components, the total (broad + narrow) H$\beta$ best-fit model is used to estimate H$\beta$ line flux. The H$\gamma$ emission line complex consists of broad and narrow H$\gamma$ and [O III] $\lambda$4363 lines. Since H$\gamma$ is much weaker than H$\beta$ the spectral decomposition is difficult to perform, especially for low S/N spectra. Therefore, instead of modeling the H$\gamma$ complex, emission line flux is directly integrated using the best-fit continuum (AGN power-law and Fe II) subtracted spectra. Furthermore, due to low S/N in some epochs and blending with [O III] $\lambda$4959, H$\beta$ line wings are not well constrained. Therefore, to minimize any systematic uncertainty in spectral decomposition, the H$\beta$ and H$\gamma$ line fluxes are integrated within $4800-4930\AA$ and $4290-4410\AA$, respectively, to avoid the line wings. 

Uncertainties in the spectral model parameters (e.g., flux, FWHM, $\alpha$) are estimated by Monte Carlo simulation. For each observed spectrum, 100 mock spectra are generated adding Gaussian random deviates of zero mean and sigma being the associated observed flux uncertainty. Then the same spectral decomposition method is repeated on the mock spectrum as done for the observed spectrum. The distribution of each parameter from the 100 mock spectra for each original spectrum allowed us to calculate 1$\sigma$ (68\%) dispersion, which was considered as the measurement uncertainty of that parameter.

The final analysis is performed on 271 spectra excluding 70 spectra with poor continuum fitting and low S/N. The final $5100\AA$, H$\beta$, and H$\gamma$ spectroscopic light curves are shown in Figure \ref{Fig:lc} and given in Table \ref{Table:data}.  The variation of the optical spectral index with time is shown. The $\gamma$-ray and radio 15 GHz light curves are also shown in Figure \ref{Fig:lc}.

    \begin{table}
    \caption{Variability statistics. Columns are (1) light curve (2) median flux of the light curve in the units of $10^{-15}$ erg s$^{-1}$ cm$^{-2}$ \AA$^{-1}$ for $f_{5100}$ and $10^{-15}$ erg s$^{-1}$ cm$^{-2}$ for emission lines, (3) fractional root-mean-square variability in percentage, (4) the ratio of maximum to minimum flux variation, (5) average cadence in days over the entire light curve and over each season in parenthesis.}
   	\begin{center}
    	\resizebox{1.0\linewidth}{!}{%
        \begin{tabular}{ l l l l l l}\hline \hline 
        light curve & median flux     &  $F_{\mathrm{var}}$ (\%)   & $R_{\mathrm{max}}$ &  Cadence  \\
                    &                 &                        &                      &  Entire (Season) \\
        (1)         & (2)             &   (3)                  & (4)                  & (5)\\ \hline
        $f_{5100}$  & 1.09 $\pm$ 0.44 & 37.30 $\pm$ 0.06       & $7.36 \pm 0.08$ & 11.4 (6.3) \\                  
        H$\beta$    & 46.96$\pm$ 6.10 & 11.88 $\pm$ 0.29       & $2.09 \pm 0.12$ & {\textemdash}\\
        H$\gamma$   & 22.92$\pm$ 3.18 & 9.61  $\pm$ 0.71       & $2.51 \pm 0.58$ & {\textemdash}\\ 
        \hline \hline
           \end{tabular} } 
           \label{Table:var}
           \end{center}
       \end{table}

\subsection{Variability}

In order to characterize the flux variation in different wavelengths, the fractional root-mean-square (rms) variability amplitude is calculated following \citet{1997ApJS..110....9R}.

\begin{equation}
F_{\mathrm{var}} = \frac{\sqrt{\sigma^2 - <\delta^2>}}{<f>},
\end{equation}
were $\sigma^2$ is the variance, $<\delta^2>$ is the mean square error, and $<f>$ is the arithmetic mean of the light curves. The ratio of maximum to minimum flux variation ($R_{\mathrm{max}}$) is also calculated for photometric and spectroscopic light curves. The values are given in Table \ref{Table:var}. The source shows strong variations in all bands from $\gamma$-ray to radio. Optical photometry and 5100$\AA$ spectroscopic light curve also show strong variation. Noteworthy is the correlation of flux variation in optical and $\gamma$-ray bands. There are two strong peaks at MJD $=$ 54900 and 57150, where the optical flux shows correlated variation with $\gamma$-ray, however, emission lines do not show any correlated peaks suggesting that PKS 1510-089 has a significant non-thermal synchrotron contribution.

To understand if the continuum variability of PKS 1510-089 is dominated by accretion disk (thermal contribution) or jet (non-thermal synchrotron contribution), the Non-Thermal Dominance parameter \citep[NTD;][]{2012Shaw} is calculated following \citet{2016FSAS}

\begin{equation}
\mathrm{NTD} = \frac{L_{o}}{L_{p}} = \frac{(L_{d} + L_{j})}{L_{p}},
\end{equation}
where $L_{o}$ and $L_{p}$ are the observed continuum luminosity and predicted disk continuum luminosity estimated from the broad emission line, respectively. The observed continuum luminosity of radio-loud sources is a combination of luminosity emitted from the accretion disk ($L_{d}$) and the jet ($L_{j}$). Therefore, if the thermal emission from the disk is only responsible for ionizing the broad line clouds then $L_{p}=L_{d}$ and NTD $=$ 1 + $L_{j}/L_{d}$. If the continuum is only due to the thermal contribution from disk, then NTD $=$ 1, however, if jet also contributes to the continuum luminosity then NTD$>1$. In the case of jet contribution greater than the disk, NTD can be larger than 2. To estimate $L_{p}$, the correlation of $L\mathrm{(H\beta)}-L_{5100}$ (orthogonal least square) obtained by \citet{2019arXiv191010395R} for non-blazers SDSS DR14 quasars and $L\mathrm{(H\beta)}$ estimated in this work is used. The variation of NTD with time is shown in the last panel of Figure \ref{Fig:lc}. Following points to be noted 1) the NTD varies between 1 to 2 most of the time suggesting that the non-thermal emission from the jet is contributed to the continuum variation but thermal disk contribution dominates over the jet in the continuum luminosity. 2) At a few instants, MJD $=$ 54900 and 57150, the NTD shows strong spikes, which are correlated with the flaring event in the $\gamma$-ray light curve, increases up to 5, and 7, respectively. 

\begin{figure}
\resizebox{8cm}{14cm}{\includegraphics{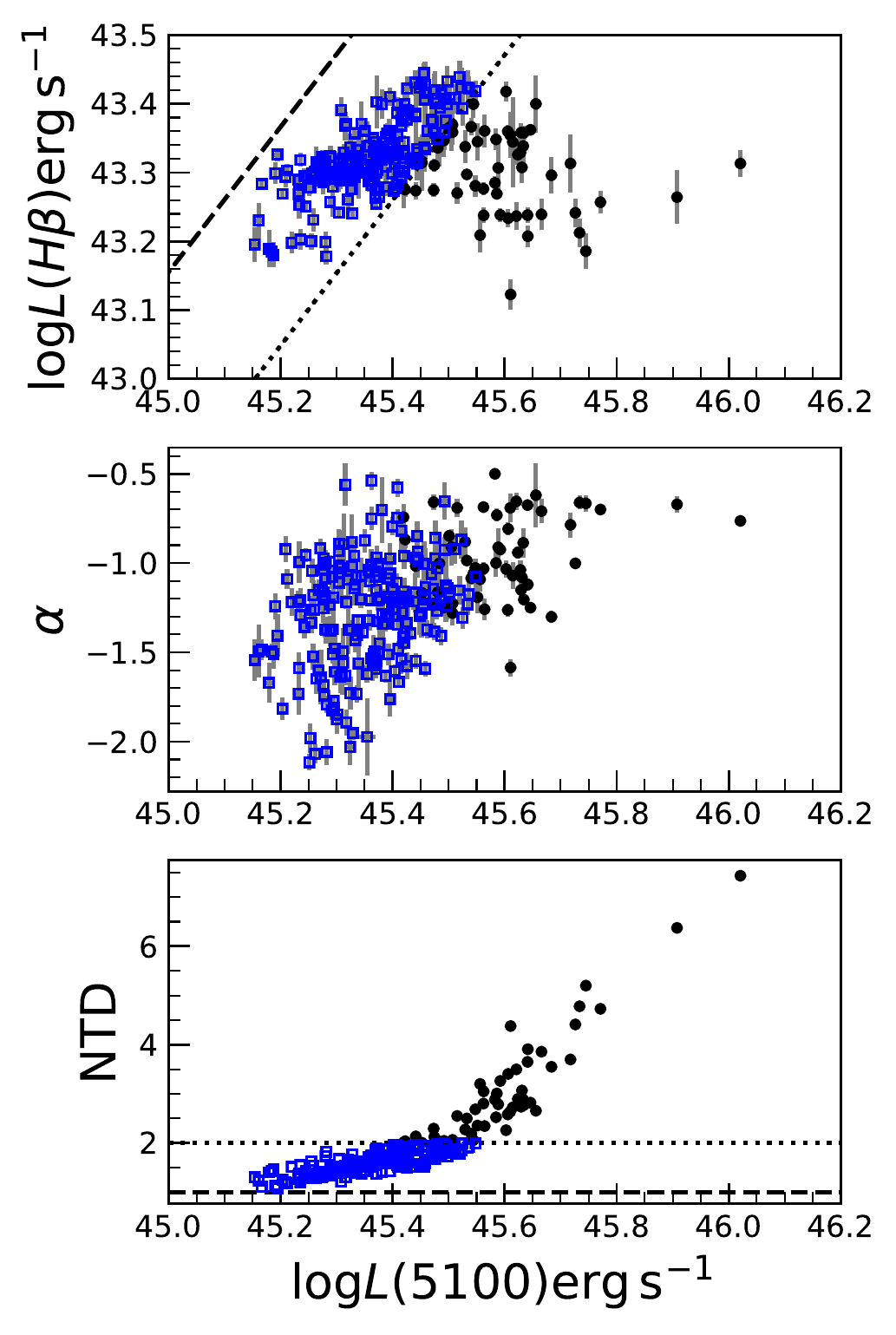}} 
\caption{Correlation of H$\beta$ line luminosity (upper panel), optical spectral index (middle panel), and NTD (bottom panel) with $L_{5100}$ during the monitoring period. The empty squares are the epochs with NTD$<$2, while filled circles are those with NTD$>=$2. The dashed and dotted lines represent NTD=1 and 2, respectively. }\label{Fig:lum} 
\end{figure}

The correlation between continuum and emission line luminosity of PKS 1510 is studied. In Figure \ref{Fig:lum}, H$\beta$ luminosity (upper panel) and NTD (bottom panel) are plotted against 5100$\AA$ continuum luminosity. The NTD gradually increases from 1 to 7 with $L_{5100}$, however, remains $<2$ until $\log L_{5100} \sim 45.6$ and increases rapidly for $\log L_{5100} > 45.6$ reaching NTD $\sim$7 for the maximum luminosity of $\log L_{5100} \sim 46$. A positive correlation between $L(\mathrm{H\beta})$ and $L_{5100}$ is found though weak with Spearman correlation coefficient ($r_s$) of 0.39 and a $p$-value of no-correlation is $10^{-11}$. This correlation becomes strong with $r_s=0.71$ ($p$-value of $10^{-35}$) when sources with NTD $<2$ (dotted line) is considered.

The spectral slope $\alpha$ ($f_{\lambda}\propto \lambda^{\alpha}$) with $L_{5100}$ is plotted in the middle panel of Figure \ref{Fig:lum}. The value of $\alpha$ increases with luminosity. For high luminosity $\log L_{5100}>45.6$, $\alpha$ is saturated and no-correlation is found with brightness. Those epochs have NTD$<$2. The median value of $\alpha$ is $-1.19\pm0.31$. A positive correlation in the $\alpha-\log L_{5100}$ relation is found with $r_s=0.43$ and $p$-value of $10^{-13}$. Therefore, a ``redder when brighter'' (RWB) trend is observed in PKS 1510 indicating the presence of accretion disk in the continuum \citep[e.g.,][]{2006A&A...450...39G,2019A&A...631A...4N}.

\begin{figure}
\resizebox{9cm}{8cm}{\includegraphics{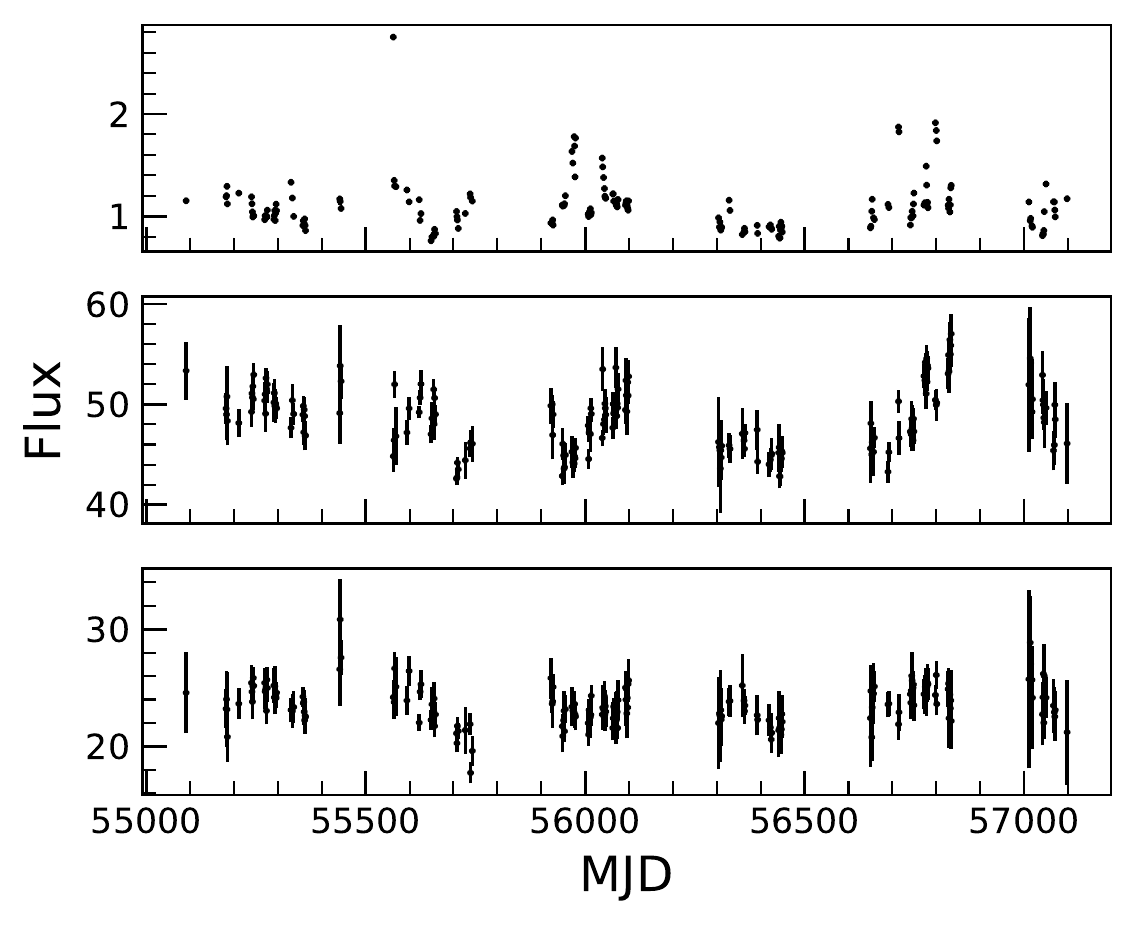}} 
\caption{The detrended spectroscopic light curves used in the time delay analysis. From top to bottom $f_{5100}$ continuum, H$\beta$ and H$\gamma$ line light curves are shown. The units are $10^{-15}$ erg s$^{-1}$ cm$^{-2}$ \AA$^{-1}$ for $f_{5100}$ and $10^{-15}$ erg s$^{-1}$ cm$^{-2}$ for line light curves.  The point above $2 \times 10^{-15}$ erg s$^{-1}$ in $f_{5100}$ light curve is excluded from time delay analysis.}\label{Fig:lc_line_detrend} 
\end{figure}

\begin{figure*}
\resizebox{9cm}{7cm}{\includegraphics{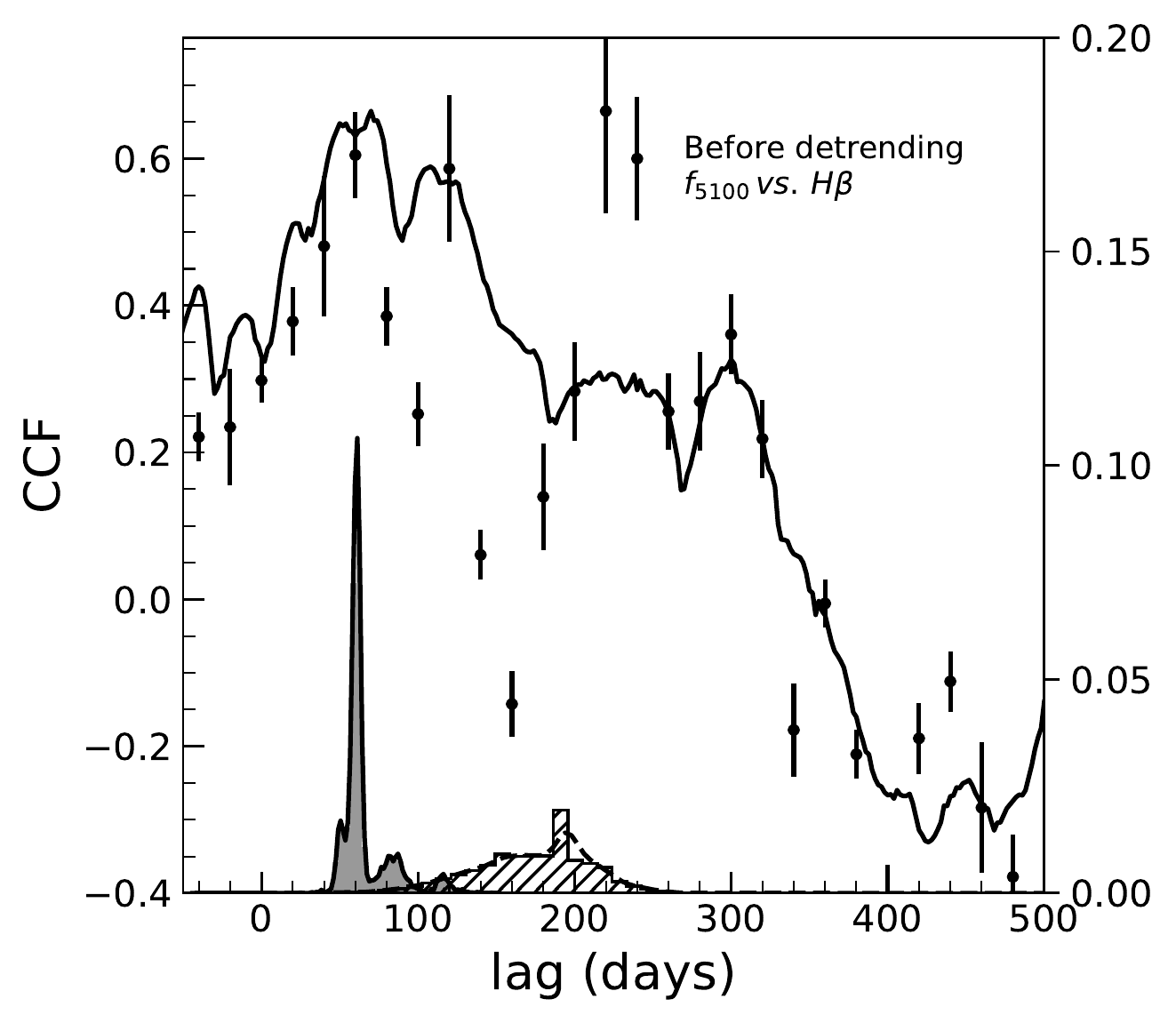}}  
\resizebox{9cm}{7cm}{\includegraphics{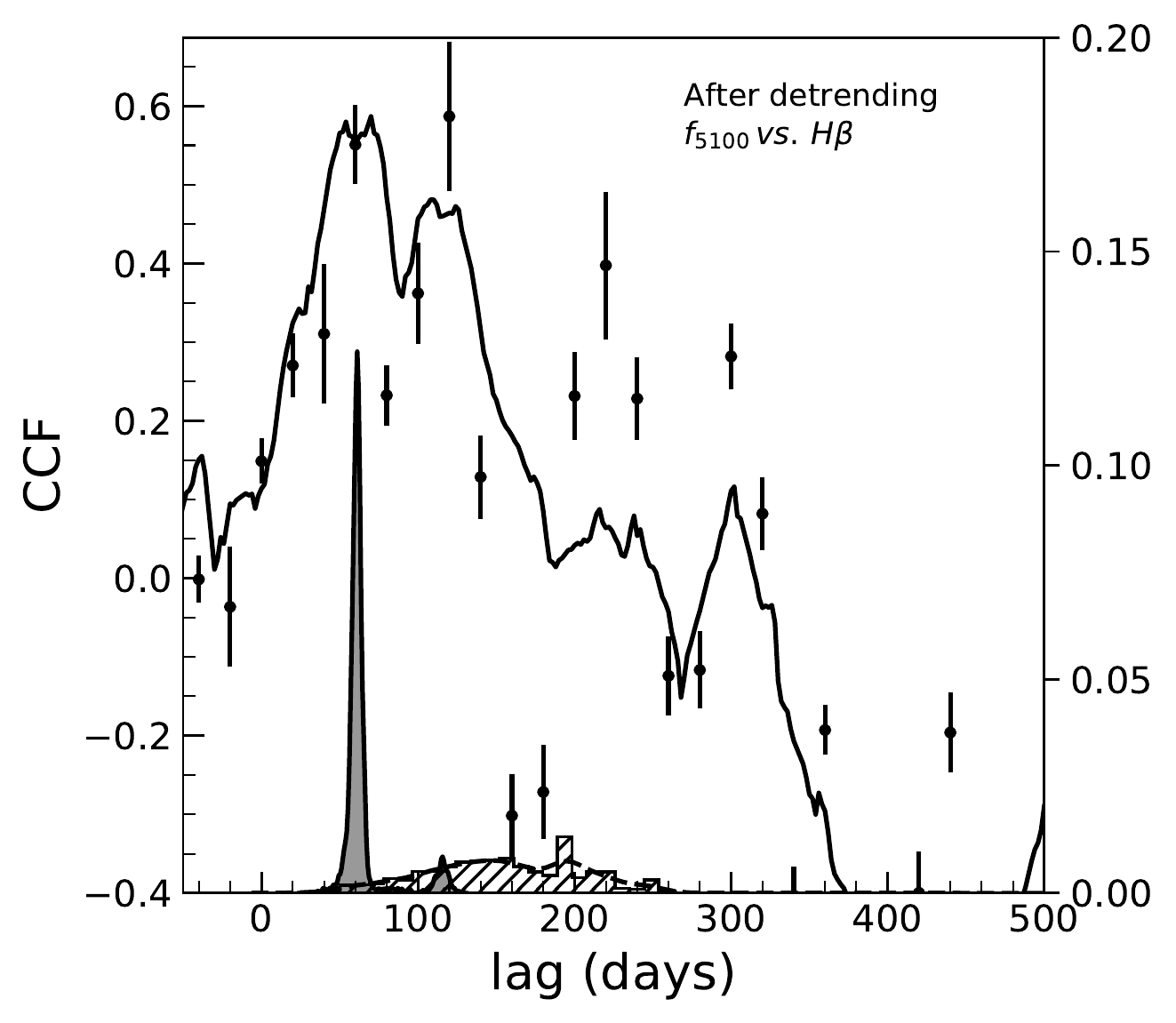}} 
\resizebox{9cm}{7cm}{\includegraphics{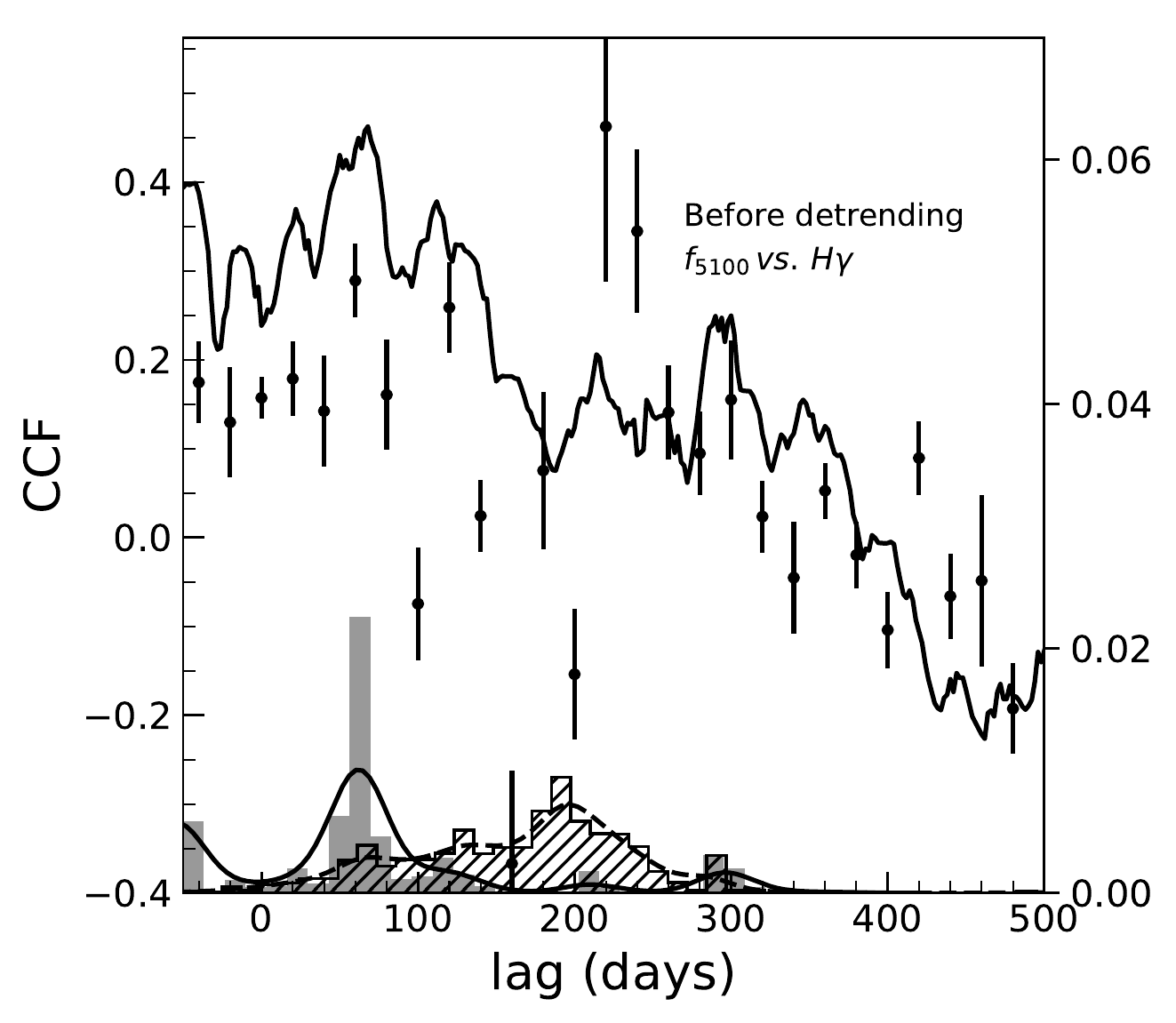}}
\resizebox{9cm}{7cm}{\includegraphics{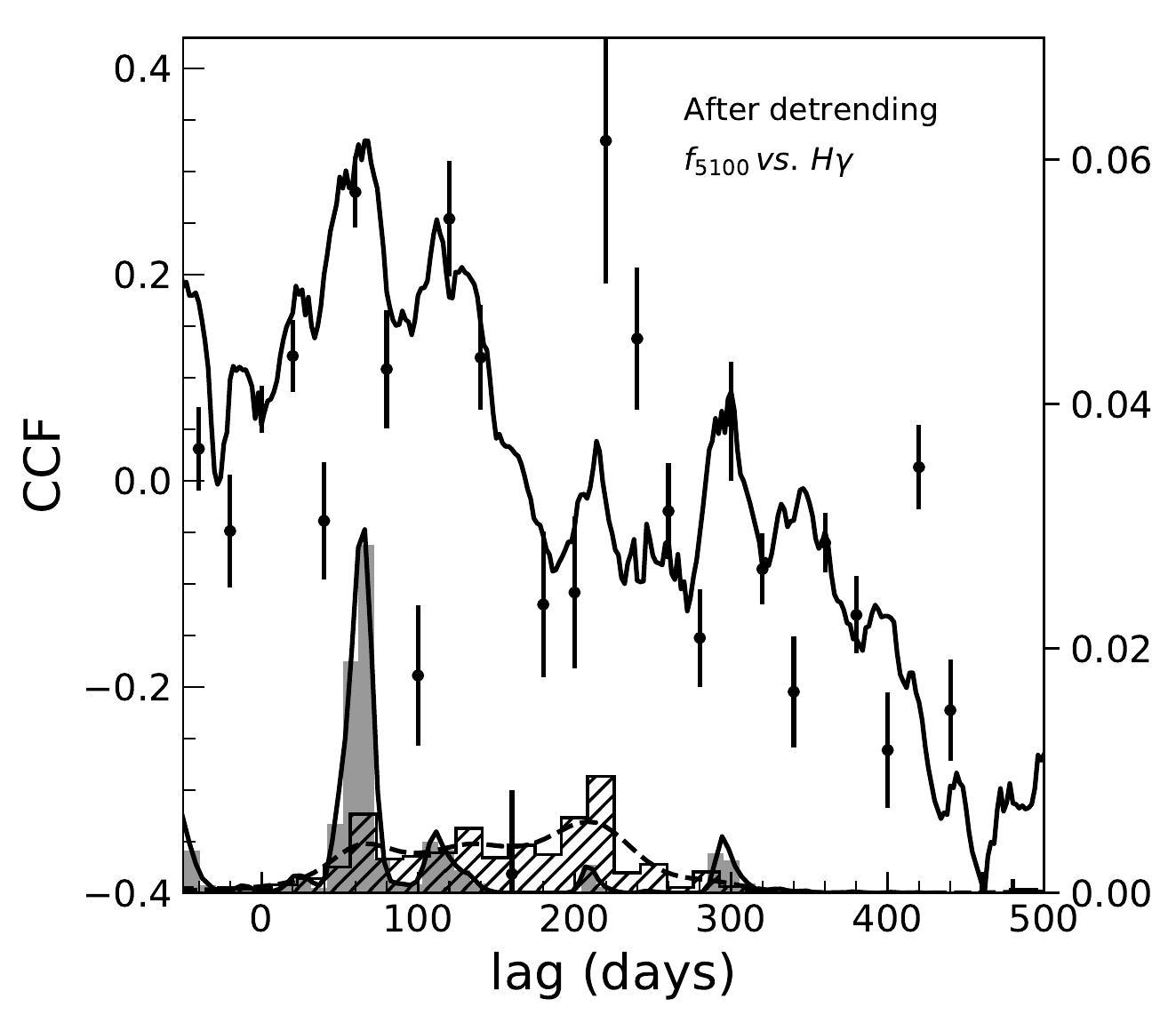}} 
\caption{Cross-correlation analysis of $f_{5100}$ vs H$\beta$ (top panels) and H$\gamma$ (bottom panels) light curves before (left panel) and after detrending (right panel). The ICCF (line) and DCF (points) are shown. The probability distribution of centroid of ICCF (filled histogram) and DCF (hatched histogram) along with smooth kernel density (solid and dashed lines, respectively) are also shown.}\label{Fig:ccf} 
\end{figure*}

\subsection{Time delay measurement}\label{sec:time_series}

As found in the previous section, the optical continuum light curve of PKS 1510-089 has a non-thermal contribution, which is dominant at some epochs where $\gamma$-ray light curve shows flare. However, broad emission line clouds do not respond to this variation. Therefore, to estimate the time delay between continuum  and emission line variation, a part of the light curve is used between MJD=55000$-$57100 denoted by the vertical line (thereafter window `A').  

The spectroscopic light curve of PKS 1510 shows a long-term trend. Such trends, which are not due to the reverberation variation \citep[see,][]{1999IOP...Welsh}, have been reported in previous reverberation mapping studies \citep[e.g.,][]{2010ApJ...721..715D,2019ApJ...876...49Z}. \citet{1999IOP...Welsh} suggested fitting a low-order (at least linear) polynomial to the light curve and subtract it from the light curve (i.e.,``detrending'') to improve the cross-correlation results. Therefore, each spectroscopic light curve is detrended prior to the cross-correlation analysis. The linear fits to the $f_{5100}$, $f\mathrm{(H\beta)}$, and $f\mathrm{(H\gamma)}$ light curves are shown in Figure \ref{Fig:lc} by the dashed line and the detrended spectroscopic light curves are shown Figure \ref{Fig:lc_line_detrend}.

\subsubsection{Cross-correlation analysis}
The cross-correlation technique \citep{1987ApJS...65....1G,1994PASP..106..879W,2004ApJ...613..682P} is used to measure the time delay. Following the description of \citet{2004ApJ...613..682P}, the interpolated cross-correlation function (ICCF) is calculated. First, the cross-correlation between the interpolated continuum light curve is performed while keeping the line light curve unchanged and calculated the cross-correlation function (CCF). Then the CCF is re-calculated with the interpolated line light curve while keeping the continuum light curve unchanged. The average of two CCFs provides the final ICCF. Additionally, the discrete correlation function (DCF) is measured following \citet{1988ApJ...333..646E}. The centroid of the CCF ($\tau_{\mathrm{cent}}$) is calculated using the points 80\% of the CCF peak.  

To estimate the uncertainty in $\tau_{\mathrm{cent}}$, the flux randomization and random subset sampling (FR/RSS) method is used \citep{1998PASP..110..660P,2004ApJ...613..682P}. This is done using Monte Carlo realizations of the light curves. First, a mock light curve is created adding Gaussian noise based on the associated flux uncertainty. Second, the same number of points are randomly selected as in the original light curve and if one epoch is selected $n$ times, the uncertainty of the flux is reduced by n$^{1/2}$. A total of 5000 mock light curves are generated and $\tau_{\mathrm{cent}}$ is estimated as done for the original light curve. The median of the $\tau_{\mathrm{cent}}$ distribution is taken as the final $\tau_{\mathrm{cent}}$ and its upper and lower uncertainty are calculated such that 15.87\% of the realizations fall above and below the range of uncertainties, respectively.

The ICCF and DCF between $f_{5100}$ and H$\beta$ (top panels) and H$\gamma$ (bottom) emission line light curves are shown in Figure \ref{Fig:ccf} before (left panel) and after detrending (right panel). In Table \ref{Table:result_ccf}, the results of the cross-correlation analysis are given. Both the ICCF and DCF methods show consistent results. First, reverberation mapping lag is clearly seen from cross-correlation analysis for both the lines before and after de-trending as the highest significant peak is at the same position remains unchanged. Second, for both the H$\beta$ and H$\gamma$ light curves, the peak of the CCF before detrending is much broader or flatter than that of after detrending although the maximum correlation coefficient ($r_{\mathrm{max}}$) is slightly higher in the former. Since the CCF after detrending is much narrower and the estimated lags are well constrained, therefore, the detrend lag measurement is adopted for further analysis.

\begin{figure}
\resizebox{9cm}{4.2cm}{\includegraphics{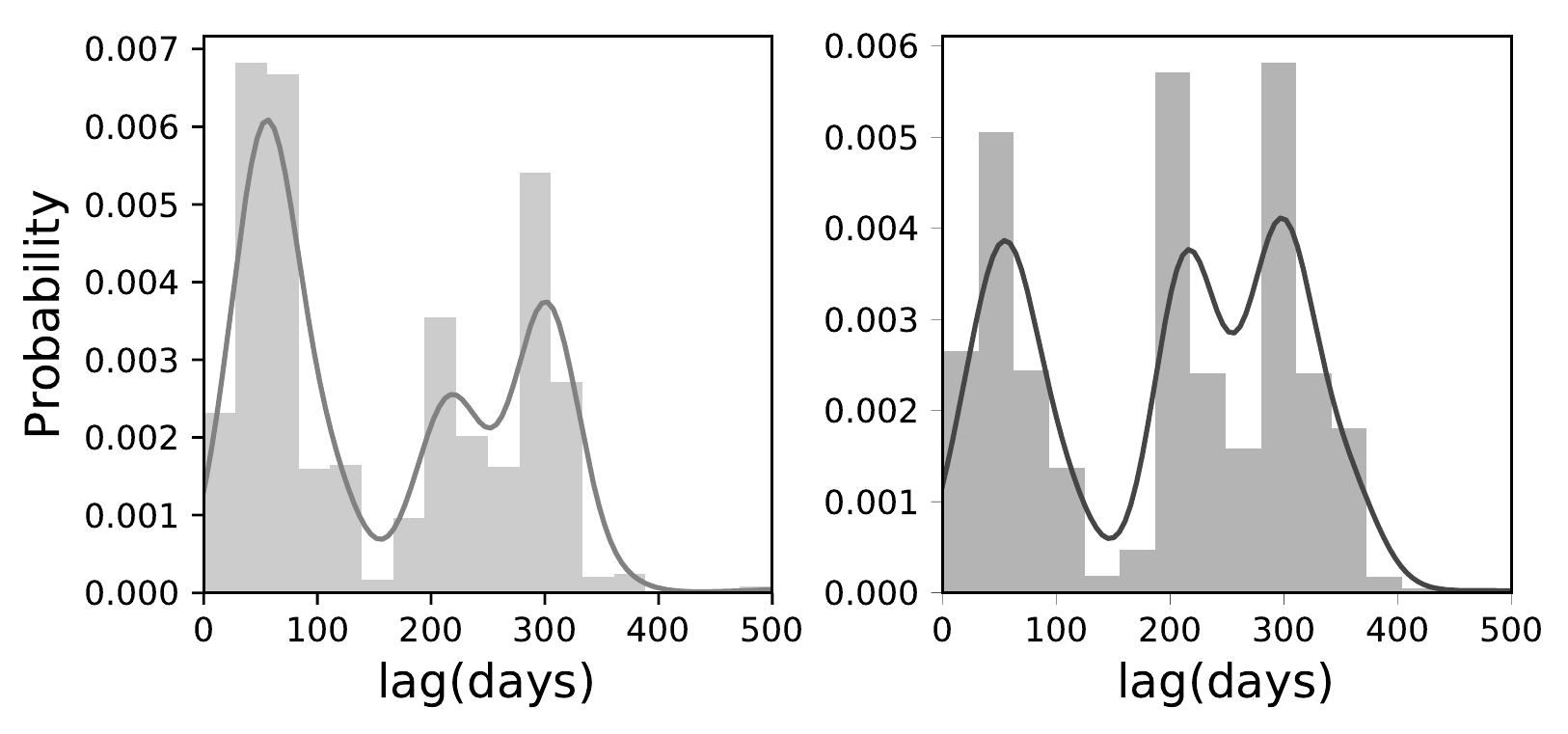}} 
\resizebox{9cm}{4.2cm}{\includegraphics{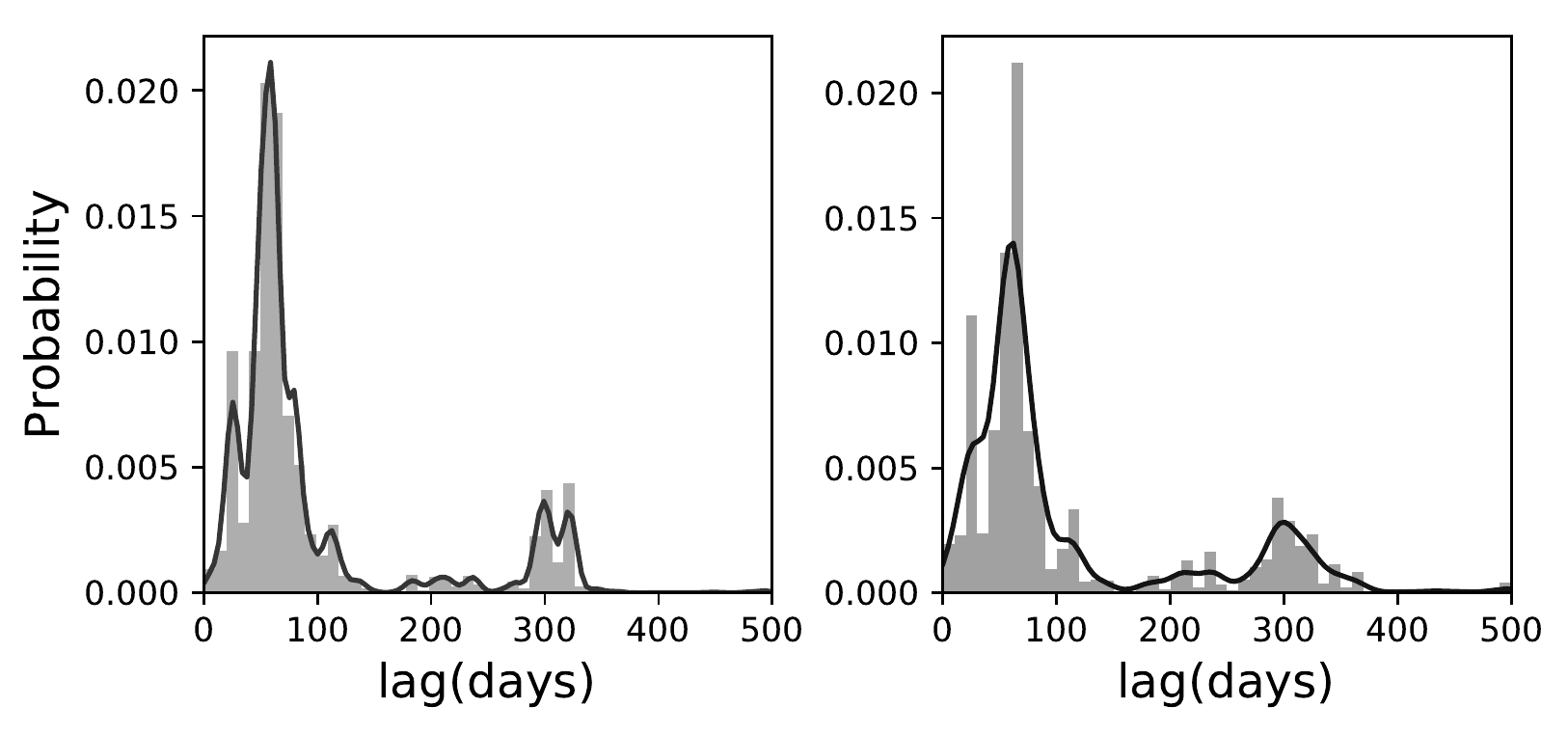}}  
\caption{Probability distribution of the observed frame time lag based on von Neumann’s estimator (top panels) and Bartels estimator (bottom panels) for H$\beta$ (left) and H$\gamma$ (right).}\label{Fig:vnrm} 
\end{figure}

\subsubsection{von Neumann and Bartels estimator}
\citet{2017ApJ...844..146C} introduced a method to measure time lag based on the regularity or randomness of data. This method does not require interpolation or binning nor the stochastic modeling of the light curves. They found that the von Neumann’s
mean-square successive-difference estimator \citep{vonneumann1941} provides better time delay measurement for irregularly sampled time series where the underlying variability process can not be modeled properly. A detailed description of this method is given in \citet{2017ApJ...844..146C}. To estimate the time delay between detrended $f_{5100}$ and line light curves of PKS 1510, a publicly available python code\footnote{\url{http://www.pozonunez.de/astro_codes/python/vnrm.py}} for optimized von Neumann’s estimator is used. The distribution of time delay obtained from von Neumann’s method after Monte Carlo simulation of FR/RSS as done for CCF analysis is shown in the upper panels of Figure \ref{Fig:vnrm}. Both H$\beta$ and H$\gamma$ show strong peaks at $\sim 60$ days, however, two additional peaks at around 200 and 300 days are also present. A modification of the von Neumann’s estimator is the Bartels estimator \citep{Bartels} that can also be used to measure time delay based on regularity or randomness of data. The distribution of time delay based on Bartels estimator is shown in the lower panels of Figure \ref{Fig:vnrm}. Unlike von Neumann’s estimator, Bartels estimator shows a single prominent peak in the distribution. The peaks at around 200 days are absent for both the H$\beta$ and H$\gamma$ light curves and the peaks at 300 days are insignificant compared to the prominent peak at around 60 days. Lag results are given in Table \ref{Table:result_ccf}.

\subsubsection{JAVELIN}\label{sec:JAVELIN}
\begin{figure}
\resizebox{9cm}{7.5cm}{\includegraphics{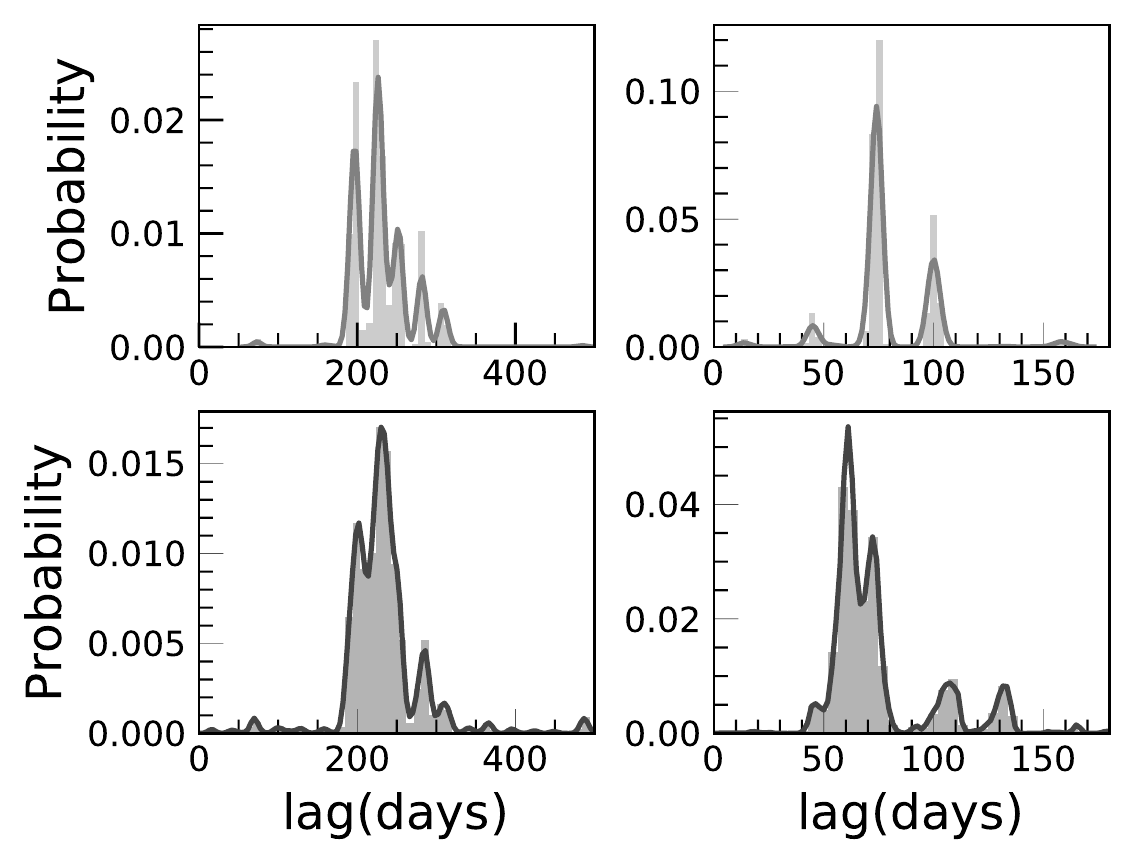}}
\caption{Probability distribution of observed frame lag computed by \textsc{javelin} when lag search is allowed between 0 to 500 (left panels) and 0 to 180 days (right panels) for H$\beta$ (upper panel) and H$\gamma$ (lower panels).}\label{Fig:javelin_lag} 
\end{figure}

 Time delay is also measured by modeling the continuum and line light curves using the \textsc{javelin} code developed by  \citet{2011ApJ...735...80Z,2013ApJ...765..106Z}. \textsc{javelin} first model driving continuum light curve by a damped random walk (DRW) process \citep[e.g.,][]{2009ApJ...698..895K} using two parameters, amplitude and time scale of variability. The emission line light curve is a shifted, scaled, and smoothed version of the continuum light curve. It then uses a Markov chain Monte Carlo (MCMC) approach to maximize the likelihood of simultaneously modeling the continuum and line light curves. In Figure \ref{Fig:javelin_lag}, the probability distribution of observed frame lag is plotted in the left panel as computed by \textsc{javelin} when a lag search is allowed between 0 to 500 days as was done for CCF and von Neumann's case. \textsc{javelin} shows prominent peaks at $\sim$200 and $\sim$ 250 days both for H$\beta$ and H$\gamma$. The peaks at around $\sim$ 60 days, which are found in the case of ICCF, von Neumann, and Bartels methods are not visible. To find any peak at lower lag, \textsc{javelin} is allowed to search lags between 0 to 180 days by refitting the light curves. The resultant lag probability distribution is shown in the right panels of Figure \ref{Fig:javelin_lag}. In this case, a prominent peak  at $\sim$ 70 days is found both for H$\beta$ and H$\gamma$ light curves. Time delays obtained from \textsc{javelin} are given in Table \ref{Table:result_ccf}.

\subsubsection{$\chi^2$-minimization}
\begin{figure}
\resizebox{9cm}{4.5cm}{\includegraphics{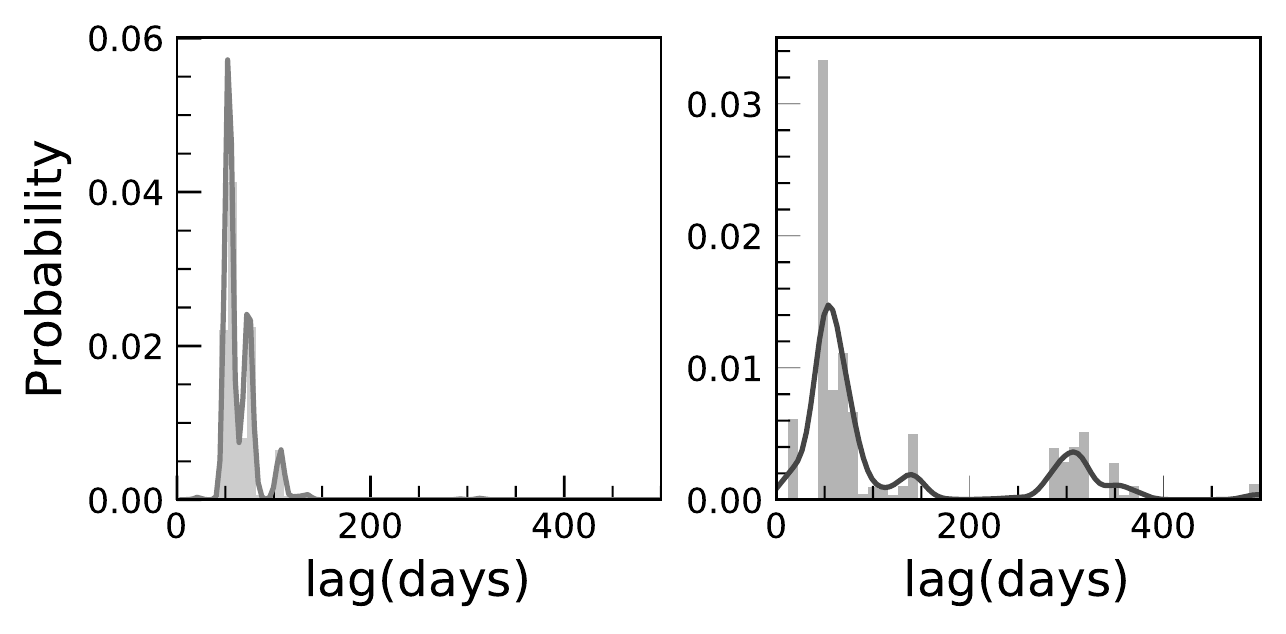}}
\caption{Probability distribution of observed frame lag computed based on $\chi^2$ minimization method.}\label{Fig:chi2_lag} 
\end{figure}

 \citet{2013A&A...556A..97C} found that $\chi^2$-minimization is a useful method to measure time lag. Therefore, to calculate time lag $\chi^2$-minimization is also applied. First, the mean values are subtracted from the light curves and then normalized by their corresponding standard deviation. The continuum light curve is then linearly interpolated to the emission line light curve and the degree of similarity is calculated by time-shifting the line light curve by $\chi^2$-minimization method. The time lag at which $\chi^2$ shows the minimum is considered as the most likely time lag. The final time lag and its uncertainty are calculated using the FR/RSS method as done in CCF analysis for 5000 iterations. The lag probability distribution obtained from $\chi^2$-minimization is shown in Figure \ref{Fig:chi2_lag} and lag values are given in Table \ref{Table:result_ccf}. The distribution shows a strong peak at $\sim$60 days both for H$\beta$ and H$\gamma$. Although a small peak at $\sim$300 days can be found for H$\gamma$, no such peak is found for H$\beta$.  

\begin{figure}
\resizebox{9cm}{5cm}{\includegraphics{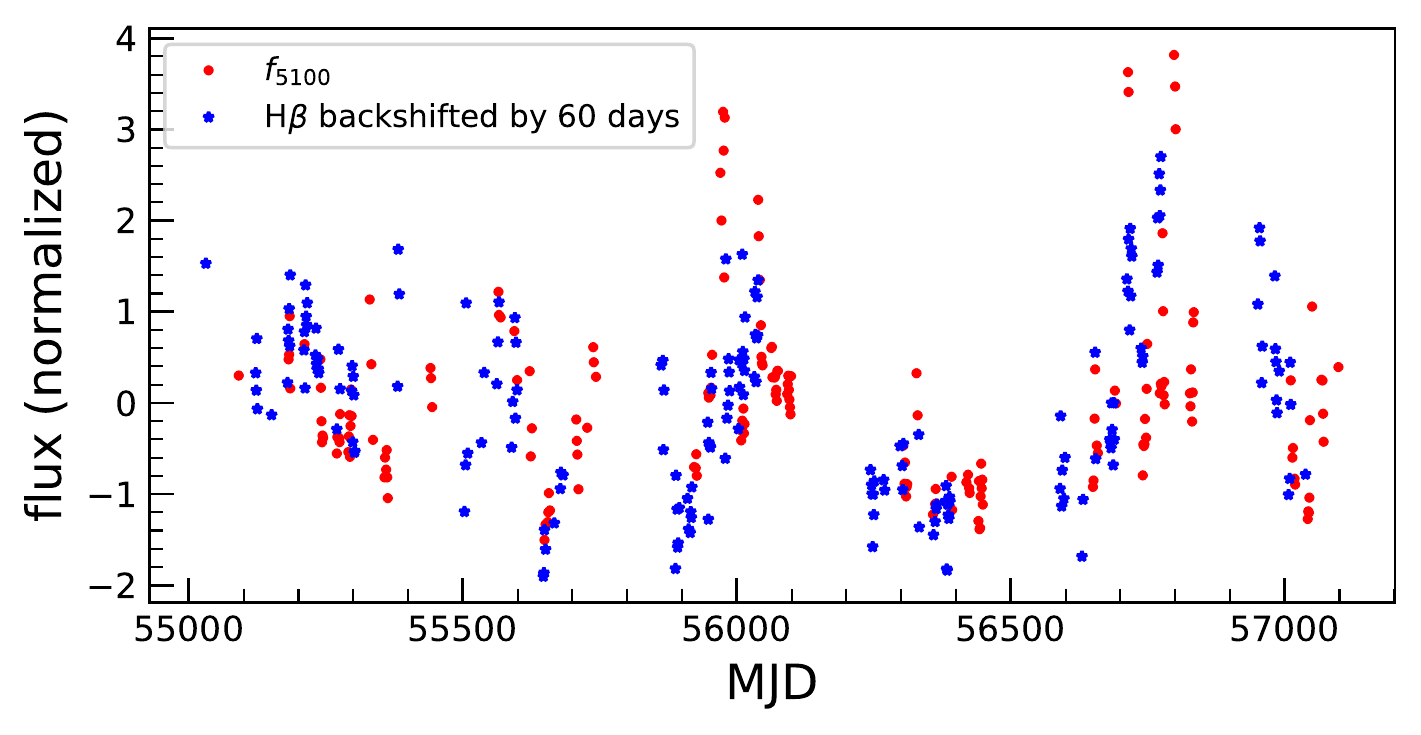}}
\caption{Normalized $f_{5100}$ light curve plotted along with the back shifted H$\beta$ light curve by 60 days.}\label{Fig:hb_backshifted} 
\end{figure}

The above methods strongly suggest a time lag of $\sim$60 days between continuum and H$\beta$ light curve. To visually check the consistency of the measured lag, in Figure \ref{Fig:hb_backshifted} we plot $f_{5100}$ light curve along with the back-shifted H$\beta$ light curve with a back-shift of 60 days. The continuum and back-shifted line light curve match well. Therefore, we adopt the lag of $61.1^{+4.0}_{-3.2}$ days obtained by the ICCF method after detrending as the best lag measurement for PKS 1510. 

 \begin{table}
 \caption{Time delay analysis results. Columns are as follows (1) method used, (2) and (3) lag for $f_{5100}$ vs. H$\beta$ and $f_{5100}$ vs. H$\gamma$ light curves. All the lags are in the observed frame.}
	\begin{center}
 	\resizebox{1.0\linewidth}{!}{%
     \begin{tabular}{ l l l }\hline \hline 
     method                &  \multicolumn{2}{c}{lag} \\\hline
                           & $f_{5100}$ vs. H$\beta$  &  $f_{5100}$ vs. H$\gamma$    \\
                           &  (days)                   &    (days)                          \\ 
      (1)                  &  (2)                      &    (3)                  \\ \hline
                           & (Before detrending)       & \\
  ICCF                     & $61.1^{+18.8}_{-4.2}$     & $63.2^{+25.8}_{-10.2}$   \\
  DCF                      & $178.8^{+29.6}_{-44.0}$   & $179.6^{+45.3}_{-99.1}$ \\ \hline
                           &  (After detrending)       &   \\ 
  ICCF                     & $61.1^{+4.0}_{-3.2}$      & $64.7^{+27.1}_{-10.6}$\\
  DCF                      & $154.1^{+50.9}_{-49.1}$   & $161.2^{+63.7}_{-96.2}$\\
  von Neumann              & $55.1^{+32.1}_{-26.3}$    & $56.6^{+31.9}_{-30.4}$\\
  Bartels                  & $58.7^{+20.8}_{-24.7}$    & $59.4^{+21.1}_{-31.7}$\\
  \textsc{javelin}\tablefootmark{a} & $226.9^{+28.1}_{-29.9}$   & $229.5^{+27.7}_{-29.7}$\\
  \textsc{javelin}\tablefootmark{b} & $74.6^{+25.1}_{-2.3}$     & $66.2^{+36.9}_{-8.0}$\\
  $\chi^2$-minimization    & $56.0^{+19.0}_{-4.0}$     & $52.0^{+22.0}_{-3.0}$  \\          
     \hline 
        \end{tabular} } 
        \tablefoot{
        \tablefoottext{a}{lag search is allowed between 0 to 500 days.}
        \tablefoottext{b}{lag search is allowed between 0 to 180 days.}
        }
        \label{Table:result_ccf}
        \end{center}
    \end{table}

\subsection{Line width and black hole mass}

\begin{figure}
\resizebox{9cm}{9cm}{\includegraphics{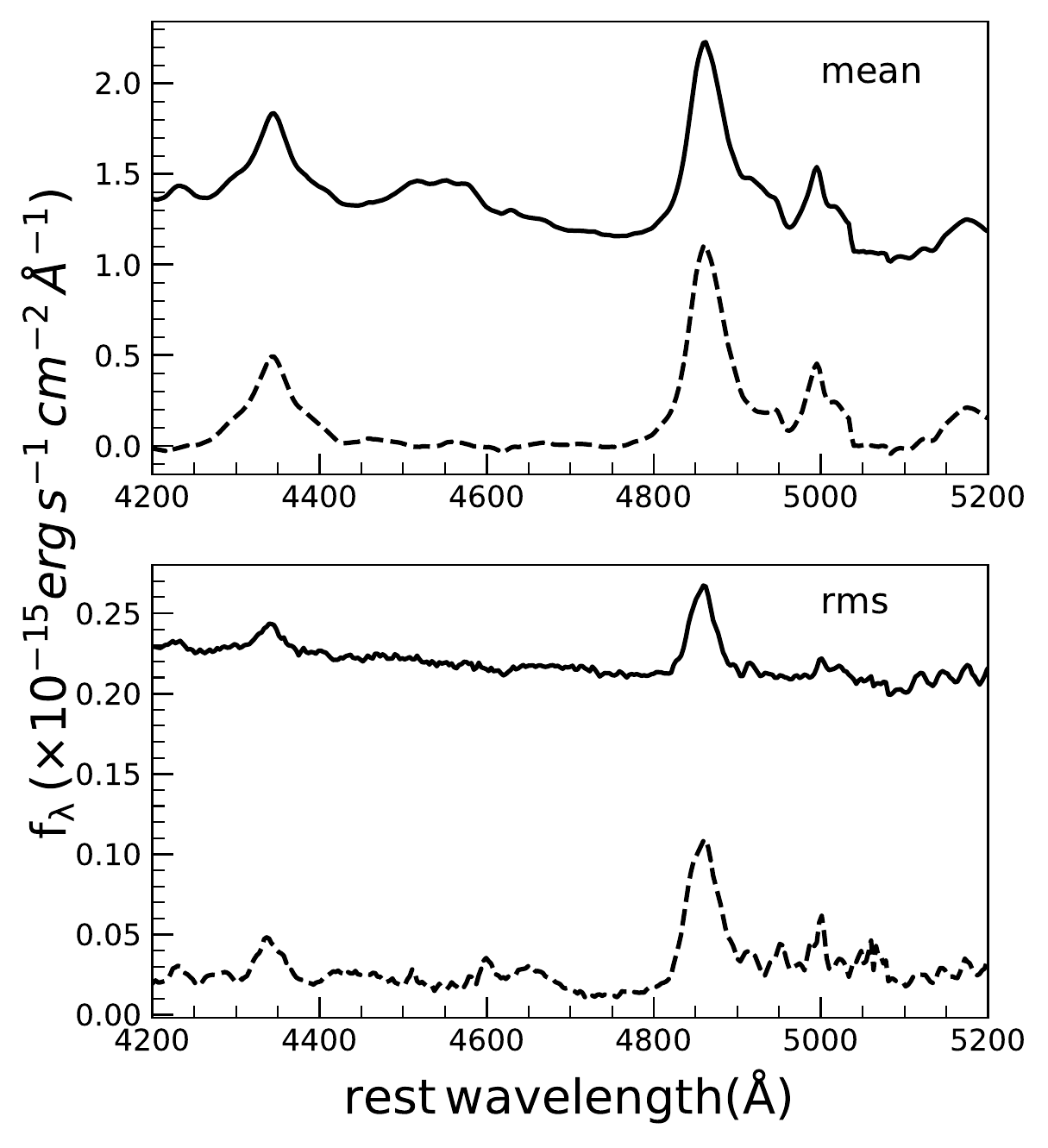}}  
\caption{The mean and rms spectra of PKS 1510-089. Top panel: The standard mean spectrum (solid) and a mean spectrum constructed after subtracting the power-law and Fe II (dashed) are shown. Bottom: Same a top panel but for the rms spectrum.}\label{Fig:mean_rms} 
\end{figure}

The mean and rms spectra are constructed from the nightly spectrum observed in window `A' following \citet{2019ApJ...886...93R}. The mean spectrum is  

\begin{equation}
<f(\lambda)> = \frac{1}{N} \sum_{i=1}^{N} f_i (\lambda).
\end{equation}
Here, $f_i (\lambda)$ is the $i$th spectrum. The rms spectrum is 
\begin{equation}
\Delta (\lambda) =\sqrt{\left[ \frac{1}{N-1} \sum_{i=1}^{N} [f_i (\lambda) -  <f(\lambda)>]^2 \right]}.
\end{equation}
 where the integration runs from 1 to the total number of spectra ($N$).
 The mean and rms spectra are shown in Figure \ref{Fig:mean_rms}. The rms spectrum clearly shows variations in both the H$\beta$ and H$\gamma$ lines.

    \begin{table}
    \caption{Rest-frame resolution corrected line width and black hole mass measurements from mean and rms spectra created from the nightly spectrum after subtracting the power-law and Fe II component. Columns are (1) spectrum type (2) line width indicator (3)  line width and (4) black hole mass.}
   	\begin{center}
    	\resizebox{0.8\linewidth}{!}{%
        \begin{tabular}{ l l l  l}\hline \hline 
        Spectrum &   type                    & $\Delta V$     & $M_{\mathrm{BH}}$   \\
                 &                           & (km s$^{-1}$)  &  ($\times 10^7 M_{\odot}$) \\
        (1)      & (2)                       & (3)            &  (4)  \\ \hline
        mean     &  FWHM                     & $2673 \pm 15$   &  $7.02^{+0.46}_{-0.36}$  \\
                 &  $\sigma_{\mathrm{line}}$ & $1378 \pm 64$   &  $7.45^{+0.59}_{-0.52}$  \\
        rms      &  FWHM                     & $2066 \pm 76$   &  $4.19^{+0.31}_{-0.26}$ \\
                 &  $\sigma_{\mathrm{line}}$ & $1207 \pm 105$  &  $5.71^{+0.62}_{-0.58}$  \\
          
        \hline \hline
           \end{tabular} } 
           \label{Table:mean_rms}
           \end{center}
       \end{table}

The FWHM and line dispersion (second moment, $\sigma_{\mathrm{line}}$) are measured from the mean and rms spectra constructed from the nightly spectrum after subtracting the power-law and Fe II component. The FWHM is calculated from the methodology described in \citet{2004ApJ...613..682P}. To measure $\sigma_{\mathrm{line}}$, first, the flux weighted line center is determined as follow:
\begin{equation}
\lambda_0 = \frac{\int \lambda f_{\lambda} d\lambda}{\int f_{\lambda}  d\lambda}
\end{equation}  
and then the line dispersion as
\begin{equation}
{\sigma^2}_{\mathrm{line}} = \frac{\int \lambda^2 f_{\lambda} d\lambda}{\int f_{\lambda}  d\lambda} - {\lambda^2}_0,
\end{equation}
where $f_{\lambda}$ is the mean or the rms spectra. The endpoints of the integrations are selected visually to be $4800-4930\AA$ for H$\beta$. To estimate uncertainty in the line width measurements Monte Carlo bootstrap method \citep{2004ApJ...613..682P} is used. For each realization, from a set of N spectra, N spectra are randomly selected without replacement and the line width is calculated from the mean and rms spectra. In each realization, the endpoints of the integration window are randomly varied within $\pm 10\AA$ from the initial selections. A total of 5000 realizations are performed providing us a distribution of FWHM and $\sigma_{\mathrm{line}}$. The median of the distribution is taken as the final line width and the standard deviation of the distribution is considered to be measurement uncertainty. The final line width measurements are given in Table \ref{Table:mean_rms} after correcting for the instrumental resolution of FWHM $\sim$1150 km s$^{-1}$.

The black hole mass of PKS 1510-089 is determined using the virial relation as follows
\begin{equation}
M_{\mathrm{BH}} = f \frac{R_{\mathrm{BLR}} \Delta V^2}{G},
\end{equation}
where $f$ is the virial factor, $R_{\mathrm{BLR}}=c\tau$ is the BLR size in the rest-frame and $\Delta V$ is the velocity width of the broad emission line. 

Using H$\beta$ lag of $\tau_{\mathrm{cent}}=61.1^{+4.0}_{-3.2}$ days, which corresponds to a rest-frame BLR size of $R_{\mathrm{BLR}}= c\tau_{\mathrm{cent}}/(1+z) =44.9^{+2.9}_{-2.3}$ light-days, the black hole mass is determined. Both the resolution-corrected FWHM and $\sigma_{\mathrm{line}}$ measured from the mean and rms spectra of H$\beta$ are used as $\Delta V$. A value of $f=4.47$ (1.12) is adopted when $\sigma_{\mathrm{line}}$ (FWHM) is considered as $\Delta V$ \citep{2015ApJ...801...38W}, however, note that FWHM is a non-linear function of $\sigma_{\mathrm{line}}$ \citep[e.g.,][]{2014SSRv..183..253P,bonta2020sloan}. Finally, four different black hole masses are determined based on the four different choices of line widths (see Table \ref{Table:mean_rms}). The black hole masses and their uncertainties are calculated based on the error propagation method and given in Table \ref{Table:mean_rms}. Note that $\sigma_{\mathrm{line}}$ is less sensitive to the line peak, and FWHM is less sensitive to the line wing, therefore, black hole masses based on the $\sigma_{\mathrm{line}}$ is widely adopted as the best mass measurement \citep[e.g.,][]{2004ApJ...613..682P,2014SSRv..183..253P}. Therefore,  $M_{\mathrm{BH}}=5.71^{+0.62}_{-0.58} \times 10^{7} M_{\odot}$ of PKS 1510-089 from $\sigma_{\mathrm{line}}$ of rms spectrum is adopted in this work.

\begin{figure}
\resizebox{9cm}{8cm}{\includegraphics{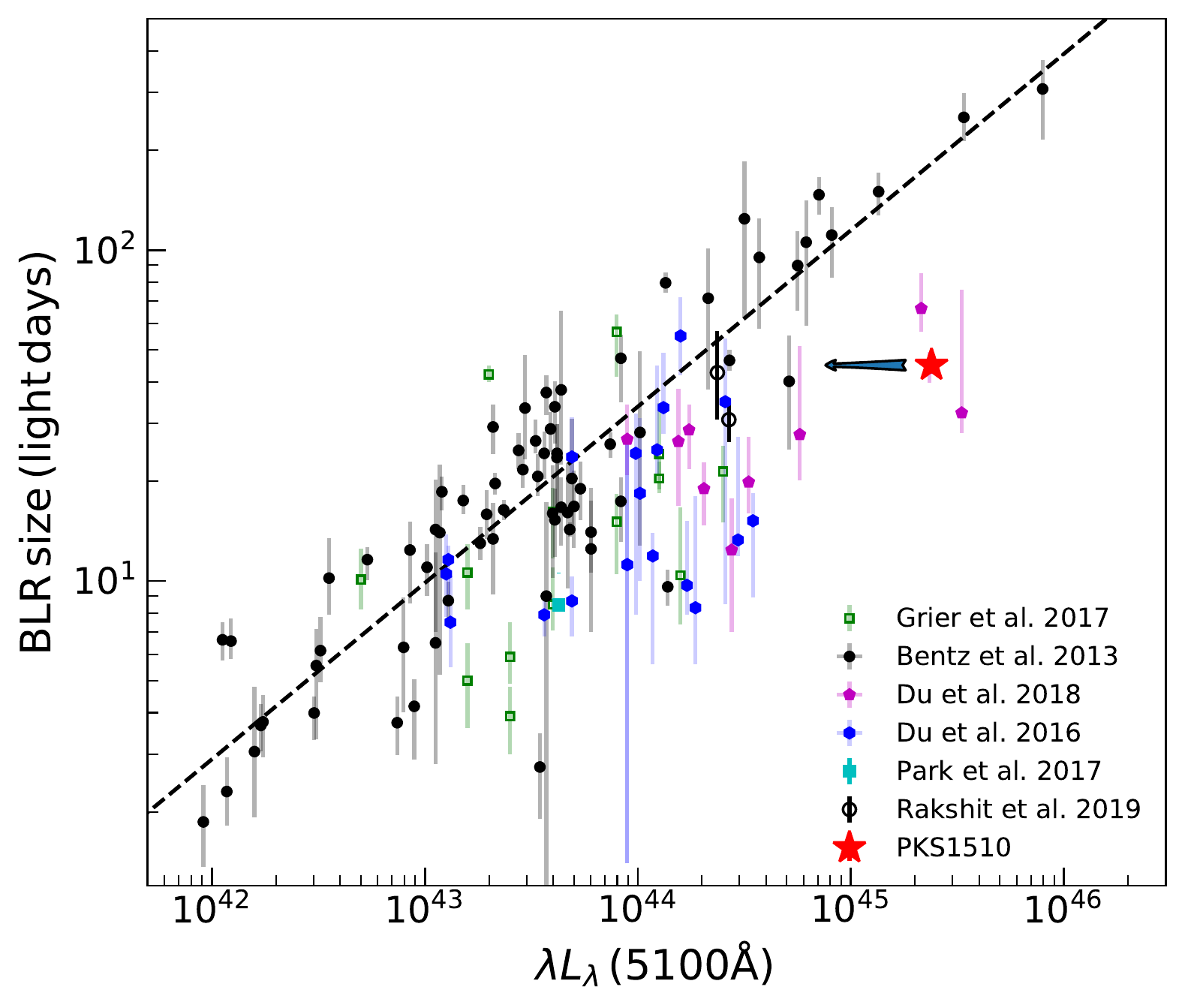}}  
\caption{PKS 1510-089 in the BLR size vs. $L_{5100}$ relation of AGNs. The best-fit relation of \citet{2013ApJ...767..149B} is shown along with various reverberation mapping results from literature. The arrow indicates that the accretion disk contribution to the measured $L_{5100}$ of PKS 1510-089 could be much lower.}\label{Fig:lag_lum} 
\end{figure}
    
\section{Discussion}\label{sec:discussion}

\subsection{Impact of seasonal gaps on lag measurement}
The monitoring of PKS 1510 is strongly affected by the seasonal gaps of about 6 months due to its low declination. Therefore, any lag close to the average seasonal gaps of $\sim$180 days, will be difficult to measure because the emission-line response to the continuum variations occurs when the source is unobservable. The measured monochromatic luminosity of PKS 1510-089 from the mean spectra is $L_{5100}=2.39\times 10^{45}$ ers s$^{-1}$. Using the $R_{\mathrm{BLR}}-L_{5100}$ relation of radio-quiet AGN sample presented by \citet{2013ApJ...767..149B} an expected BLR size of $\sim$182 days and an observed frame lag of 248 days are obtained for PKS 1510. As the expected lag is closer to the seasonal gaps the impact of seasonal gaps on the lag measurement is investigated constructing mock light curves. For this purpose, a mock continuum light curve is constructed using the DRW model implemented in \textsc{javelin} having a similar characteristic. Then mock line light curves are constructed with a time delay of 70 and 200 days. To mimic the observed light curves, mock light curves are down-sampled to have the same time sampling, therefore the same time axis, as observed light curves. To recover the input time delay, time series analysis methods as described in section \ref{sec:time_series} are used on the mock data sets. 

The results are shown in Appendix Figure \ref{Fig:mock_ccf} and are given in Table \ref{Table:result_mock}. The results show ICCF, DCF, von Neumann and Bartels recover well an input lag of 70 days and not affected by the seasonal gaps and time sampling. Note that the $r_{\mathrm{max}}$ obtained by ICCF and DCF is $\sim$0.6, which is similar to that obtained in the case of the observed light curve (see Figure \ref{Fig:ccf}). Interestingly, \textsc{javelin} results are affected by seasonal gaps as it shows a primary peak at $\sim$ 160 days. Due to the same reason, \textsc{javelin} found a lag at $\sim$ 200 days for the observed light curve of PKS 1510 when the lag search is allowed between $0-500$ days (see section \ref{sec:JAVELIN}). A secondary peak at $\sim$70 days, which is the same as the input lag, is also found in \textsc{javelin} lag probability distribution in Figure \ref{Fig:mock_ccf}. For an input lag of 200 days, both the ICCF and DCF show no correlation between mock continuum and line light curve due to the seasonal gaps, therefore, they do not allow to estimate lag. However, von Neumann, Bartels and \textsc{javelin} successfully recover the input time lag of 200 days albeit with larger uncertainty. Therefore, the uses of various time series analysis methods allow us to recover a lag of $\sim$200 days although the light curves are affected by the seasonal gaps and a lag $\sim$70 days can be well-constrained with lower uncertainty. The above simulations suggest any lag close to $\sim$ 200 days is unlikely for PKS 1510 while a lag of $\sim$ 60 days is most likely for PKS 1510.

\subsection{Size-luminosity relation}
PKS 1510-089 is a radio-loud source with strong $\gamma$-ray activity. From Figure \ref{Fig:lum}, it is clear that the measured $L_{5100}$ is a combination of non-thermal synchrotron emission from the jet and thermal emission from the accretion disk. Hence, $L_{5100}$ measurement is strongly affected by the non-thermal emission. In Figure \ref{Fig:lag_lum}, the BLR size of PKS 1510-089 is plotted against $L_{5100}$ along with the previous reverberation mapped objects from literature \citep[e.g.,][]{2013ApJ...767..149B,2016ApJ...820...27D,2017ApJ...851...21G,2018ApJ...866..133D,2019ApJ...886...93R}. PKS 1510-089 is found to deviate from the $R_{\mathrm{BLR}}-L_{5100}$ relation of \citet{2013ApJ...767..149B}. However, this is not surprising considering the fact that previous studies of high-accreting and strong Fe II emitting AGNs show significant deviation from the size-luminosity relation \citep{2016ApJ...825..126D,2018ApJ...856....6D}. This could be due to the complex radiation field and BLR geometry in high accreting AGNs, which may have slim accretion disks. The strong self-shadowing effects on the slim accretion disks may produce highly anisotropic radiation field, which depending on the accretion rate may lead to the two dynamically distinct regions of the BLR \citep{Wang_2014}. \citet{2019ApJ...886...42D} found that the $R_{\mathrm{Fe II}}$, i.e. the flux ratio of Fe II to H$\beta$, is the main driver of the shortened lag obtained in the high-accreting AGNs. They provided a new scaling relation which includes $R_{\mathrm{Fe II}}$. Using the mean $R_{\mathrm{Fe II}}$ of $0.52\pm0.09$ found in PKS 1510-089, the expected lag is about $\sim$122 days, which is lower than what is expected from the \citet{2013ApJ...767..149B} relation but still a factor 3 larger than the measured lag. Several authors \citep{1997MNRAS.286..415C,2010ApJ...721.1425A,2012ApJ...760...69N} have estimated disk bolometric luminosity ($L_{\mathrm{disk}}$) of PKS 1510-089 which is in the range of $3-7\times10^{45}$ erg s$^{-1}$. This based on a simple scaling relation $R_{\mathrm{BLR}}=10^{17} \sqrt{L_{\mathrm{disk}}/10^{45}}$ cm \citep{2009MNRAS.397..985G} provides $R_{\mathrm{BLR}}=66-102$ light-days. Our estimated rest-frame $R_{\mathrm{BLR}}$ is slightly lower than the above values.

\citet{2020ApJ...897...18L} studied a well-known FSRQ 3C273 and found that the optical continuum has two components of emissions, one from the accretion disk and another from the jet. The jet contribution is found to be 10-40\% to the total optical emissions. \citet{2001MNRAS.323..718W} showed that the synchrotron radiation from the jet contributes to the optical band, thereby increasing the total optical continuum flux. However, due to beaming, this synchrotron component does not contribute to ionizing the emission line clouds. Figure \ref{Fig:lc} shows strong $\gamma$-ray activity in PKS 1510-089 at MJD $\simeq$54900 and 57150. Although the light curve when PKS 1510-089 is mostly in a quiescent state is analyzed in this work, a non-thermal contribution from the jet is always present. To have a rough estimation of thermal contribution in $L_{5100}$, the median of the NTD is calculated in the quiescent state (window A). It is found to be $1.67^{+0.31}_{-0.23}$ indicating that the disk contribution to the measured $L_{5100}$ is about 60\%.  

This contribution can roughly be estimated from the broad-band SED also. \citet{2019ApJ...883..137P} studied broad-band SED of PKS 1510-089 in several active and quiescent state during 2015 (MJD $=$ 57000$-$57350). They modeled broad-band SED using a time-dependent two-zone emission model and estimated the synchrotron, synchrotron self-Compton (SSC), and inverse-Compton emission. The disk contribution at 5100\AA, to the total flux during one quiescent state \citep[Q2 between MJD $=$ 57180$-$57208; see Figure 3 of][]{2019ApJ...883..137P} is estimated. The disk contribution is found to be $\sim$44\% to the total flux at 5100\AA, lower than that estimated from NTD but consistent within error. If the measured $L_{5100}$ is corrected assuming the disk contribution is 60\% (based on NTD calculation), the corrected $L_{5100}$ is $1.43\times10^{45}$ ers s$^{-1}$. In fact, using the mean H$\beta$ line luminosity of $2.04\times10^{43}$ erg s$^{-1}$ and the scaling relation of $L\mathrm{(H\beta)}-L(5100)$ for radio quiet AGNs from \citet{2006ApJ...637..669L},  $L_{5100}$ is found to be $1.40 \times10^{45}$ ers s$^{-1}$. Therefore the measured $L_{5100}$ of PKS 1510-089 is an upper-limit as it is significantly affected by the non-thermal contribution from the jet. Furthermore, the host galaxy also contributes to the measured $L_{5100}$. Therefore, the measured $L_{5100}$ is an upper limit and the actual position of PKS 1510-089 in the size-luminosity diagram is highly uncertain.

\subsection{Black hole mass measurement}
Since PKS 1510-089 is a well-studied object, several authors have reported its black hole mass based on single-epoch spectrum, variability time scale, accretion disk modeling, etc. Here, the mass measurement is compared with those in the literature. The estimated black hole of PKS 1510-089 ranges from $4.19-7.02 \times 10^7 M_{\odot}$ depending on the choice of the line width and line profile. The bolometric luminosity is found to be $21.51 \times 10^{45}$ erg s$^{-1}$ based on the measured $L_{5100}$ from the mean spectrum using $L_{\mathrm{BOL}} = 9 \times L_{5100}$ \citep{2000ApJ...533..631K}. The Eddington ratio ($\lambda_{\mathrm{EDD}}$) estimated using $L_{\mathrm{EDD}}= 1.26\times 10^{38} M_{\mathrm{BH}}$ is 2.98 considering the black hole based on the $\sigma_{\mathrm{line}}$ of rms spectrum as the best measurement. However, using the corrected $L_{5100}$ due to jet contribution, $\lambda_{\mathrm{EDD}} =1.78$ is found suggesting that PKS 1510-089 is accreting at a mildly super-Eddington rate. This value is similar to $\lambda_{\mathrm{EDD}}=2.4$ found for another FSRQ 3C273 \citep{2019ApJ...876...49Z}.

\citet{Oshlack_2002} using single-epoch spectrum estimated black hole mass of PKS 1510-089 as $3.86\times 10^{8} M_{\odot}$. Similarly, \citet{2005AJ....130.2506X} based on the minimum time scale of variability and single-epoch spectrum estimated black hole masses of $M_{\mathrm{BH}}=1.1\times10^8 M_{\odot}$ and $1.6\times10^8 M_{\odot}$, respectively \citep[see also][]{2006ApJ...637..669L,2017ApJ...834..157P}. Note that single-epoch black hole masses depend on the choice of the scaling relation which shows significant scatter, especially highly accreting sources show larger offset from the $R_{\mathrm{BLR}}-L_{5100}$ relation. The choice of the line width measurement, FWHM or $\sigma_{\mathrm{line}}$, also affect the single-epoch mass measurement. Reverberation mapping studies of AGNs in which multiple emission lines have been observed suggest $\sigma_{\mathrm{line}}$ is a better measure for black hole mass than FWHM \citep{2004ApJ...613..682P}. Moreover, rms profile that can only be constructed from multi-epoch spectra, should be used to estimate black hole mass as it shows the variable component of the emission line, contaminating features such as constant host-galaxy contribution and narrow-line components which are present in the single-epoch spectrum or in the mean spectrum disappear in the rms spectrum \citep[see][]{2014SSRv..183..253P}. Using UV data \citet{2010ApJ...721.1425A} estimated a black hole mass of PKS 1510-089 as $M_{\mathrm{BH}}=5.4\times10^8 M_{\odot}$. On the other hand, \citet{2017A&A...601A..30C} using the \citet{1973A&A....24..337S} model estimated a black hole mass of $2.4\times10^{8} M_{\odot}$. Therefore, a large range of black hole mass, $1-9 \times 10^{8} M_{\odot}$ of PKS 1510-089 has been reported in the literature. Considering the typical uncertainty $\sim$0.5 dex in the mass measurement \citep{2013BASI...41...61S}, the estimated black hole mass of PKS 1510-089 is smaller by a factor of $2-4$ than the values reported in the literature.

\section{Conclusion}\label{sec:conclusion}
The optical spectroscopic reverberation mapping results of PKS 1510-089 are presented from $\sim$8.5-years long monitoring campaign carried out at Steward Observatory from December 2008 to June 2017. The nightly spectrum shows presence of broad H$\beta$, H$\gamma$ and Fe II emission overlaying on a blue continuum. During the monitoring program, both the optical continuum and H$\beta$ line show strong variation with fractional root-mean-square variation ($F_{\mathrm{var}}$) of $37.30\pm0.06$\% for $f_{5100}$ and $11.88\pm0.29$\% for H$\beta$ and $9.61\pm0.71$\% for H$\gamma$ light curves. With the increase of $L_{5100}$ from $10^{45.2}$ to $10^{45.6}$ erg s$^{-1}$, the H$\beta$ line luminosity increases from $10^{43.2}$ to $10^{43.5}$ erg s$^{-1}$ but decreases as $L_{5100}>10^{45.6}$ erg s$^{-1}$. Although the optical continuum is dominated by thermal radiation from the accretion disk, non-thermal Synchrotron contribution from the jet is clearly present, which does not contribute in ionizing the emission line clouds. From cross-correlation analysis, the H$\beta$ and H$\gamma$ lags are found to be $61.1^{+4.0}_{-3.2}$ days and $64.7^{+27.1}_{-10.6}$ days, respectively. This corresponds to a rest-frame BLR size of $44.9^{+2.9}_{-2.3}$ light-days for H$\beta$ and $47.5^{+19.9}_{-7.8}$ light-days for H$\gamma$. Using a scale factor of 4.47 and the $\sigma_{\mathrm{line}}$ from rms spectrum which is constructed from the nightly spectrum after subtracting the power-law and Fe II, the black hole mass of PKS 1510-089 is found to be $M_{\mathrm{BH}}=5.71^{+0.62}_{-0.58} \times 10^{7} M_{\odot}$.

\begin{acknowledgements}

Many thanks go to the referee for comments and suggestions that helped to improve the quality of the manuscript. Thanks to Prince Raj for providing their SED model component of PKS1510-089 and Neha Sharma for carefully reading the manuscript. This publication makes use of data products from the {\it Fermi} Gamma-ray Space Telescope and
accessed from the Fermi Science Support Center\footnote{https://fermi.gsfc.nasa.gov/ssc/data/access/}. Data from the Steward Observatory spectropolarimetric monitoring project were used. This program is supported by Fermi Guest Investigator grants NNX08AW56G, NNX09AU10G, NNX12AO93G, and NNX15AU81G.
This research has made use of data from the OVRO 40-m monitoring program \citep{2011ApJS..194...29R} which is supported in part by NASA grants NNX08AW31G, NNX11A043G, and NNX14AQ89G and NSF grants AST-0808050 and AST-1109911
\end{acknowledgements}

\bibliographystyle{aa}
\bibliography{ref}

\begin{appendix}\label{sec:appendix}
\section{Time series analysis of simulated light curves}

\begin{figure}
\resizebox{4.3cm}{4cm}{\includegraphics{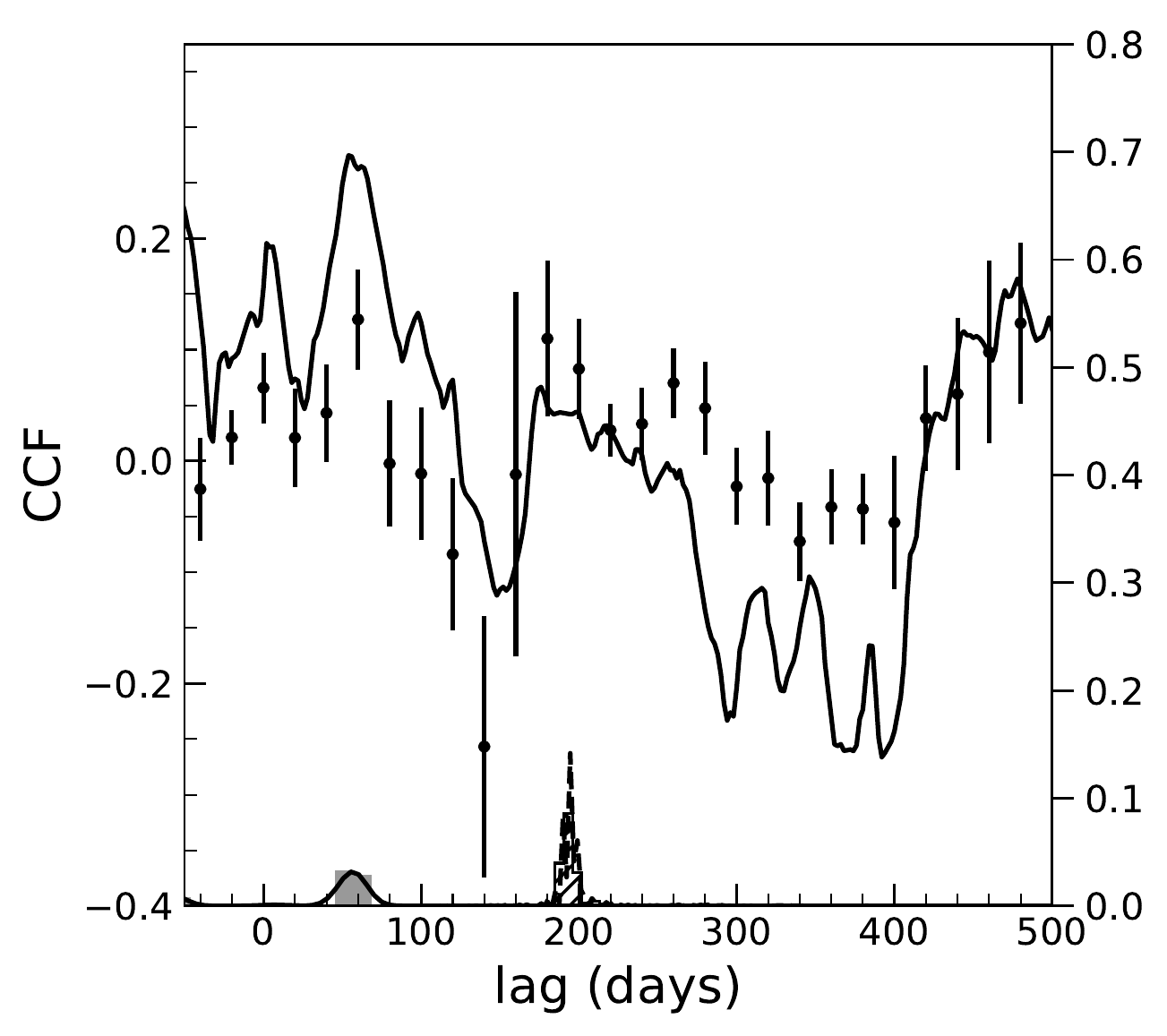}}
\resizebox{4.3cm}{4cm}{\includegraphics{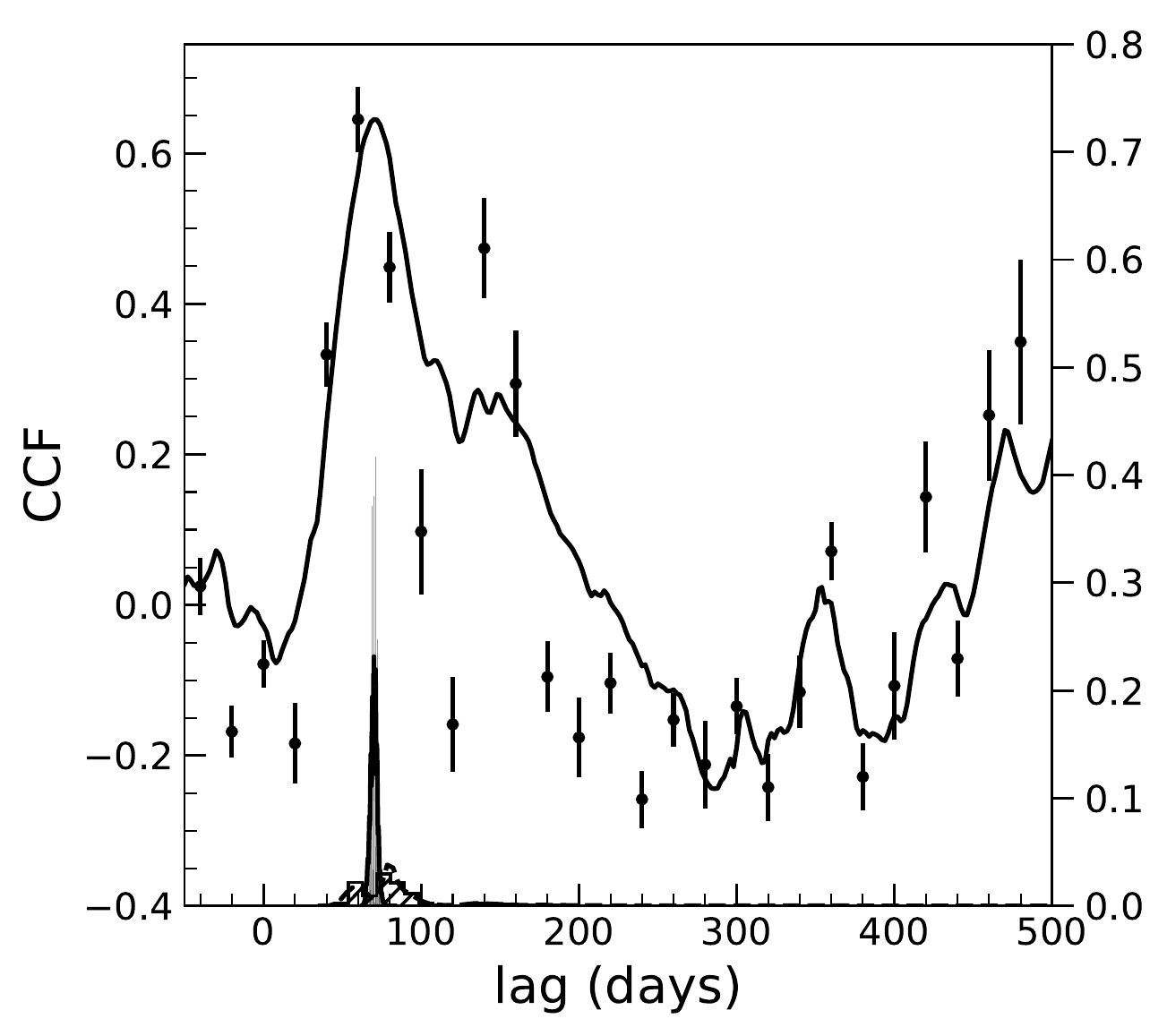}}
\resizebox{9cm}{4cm}{\includegraphics{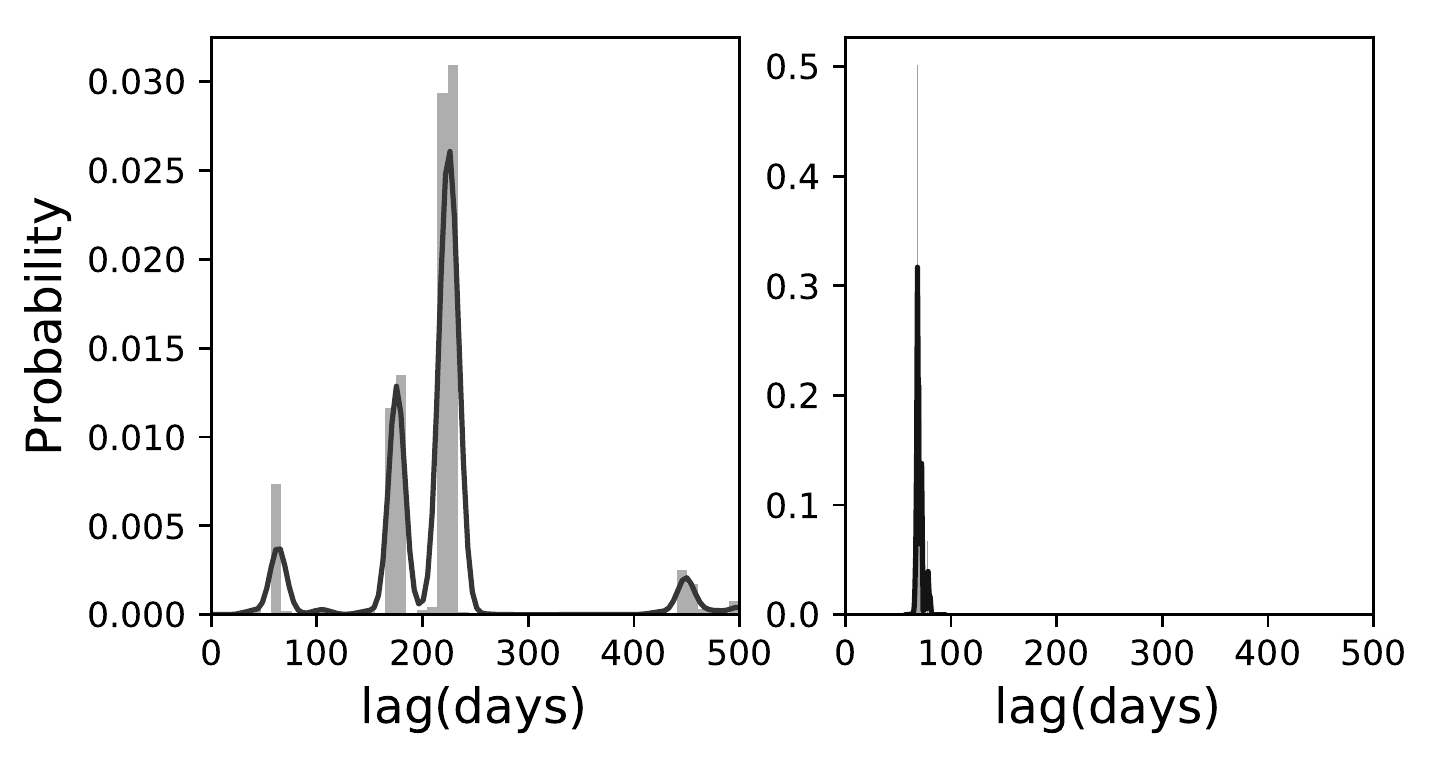}}
\resizebox{9cm}{4cm}{\includegraphics{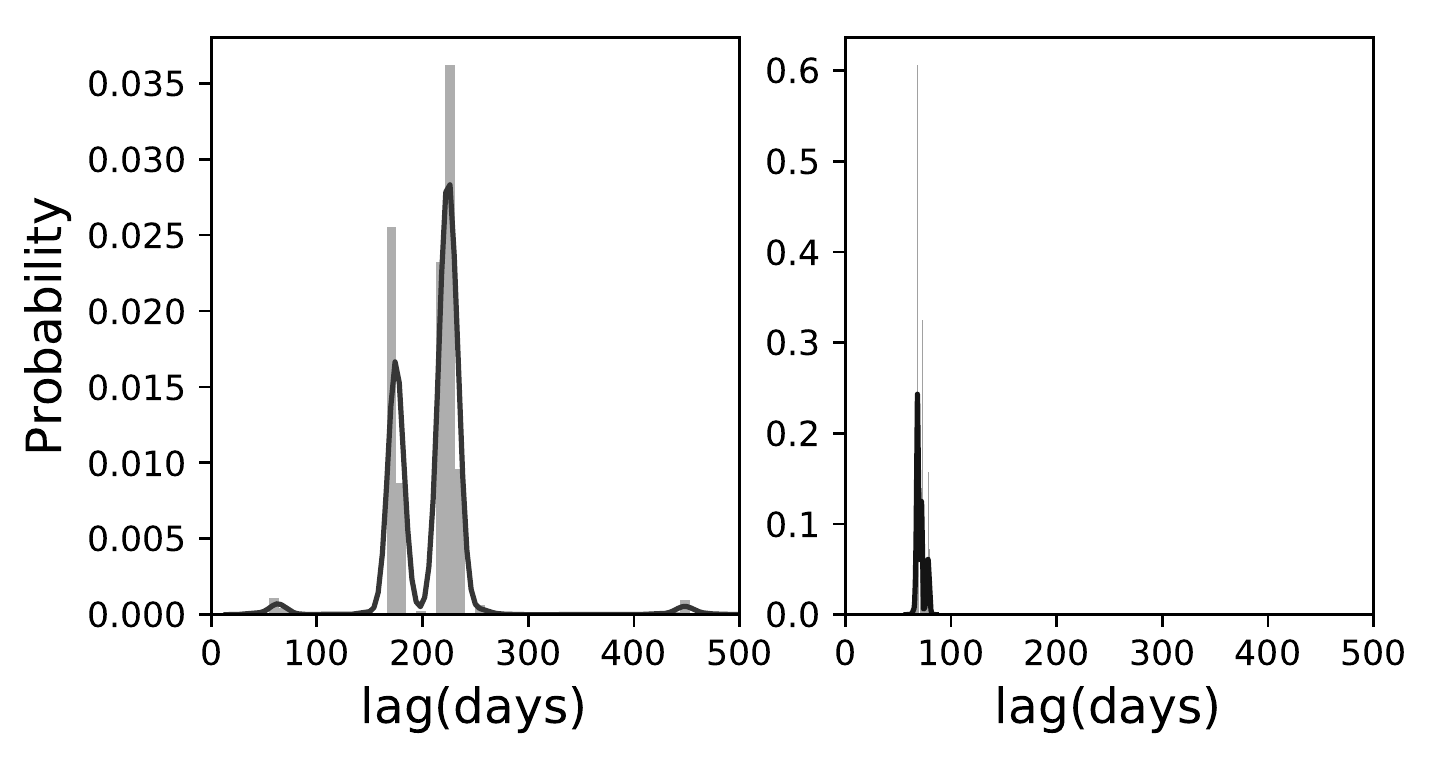}}
\resizebox{9cm}{4cm}{\includegraphics{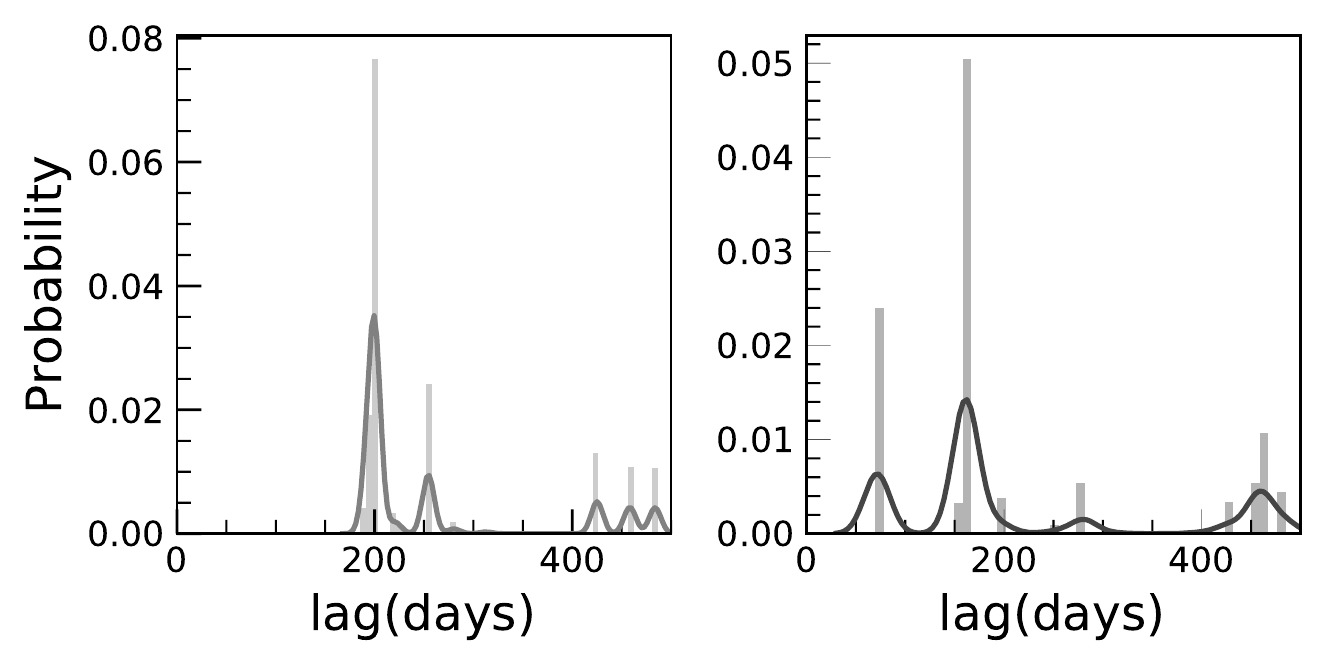}}  
\caption{From top to bottom the lag probability distributions obtained from CCF (ICCF and DCF), von Neumann's, Bartels and \textsc{javelin} are shown for mock continuum vs. mock line light curve with a delay of 200 days (left) and 70 days (right). The lag probability distribution (histogram) along with the smoothed kernel density distribution is shown in each panel.}\label{Fig:mock_ccf} 
\end{figure}

To study the impact of seasonal gaps on the lag measurement, the DRW model implemented in \textsc{javelin} is used to construct a mock continuum light curve having similar characteristics as the observed $f_{5100}$ light curve of PKS 1510. Then two mock line light curves are constructed with a lag of 70 and 200 days. The light curves are down-sampled to have the same time sampling as the observed light curves, therefore mimicking the observed time axis. The time series analysis methods (ICCF, DCF, von Neumann, Bartels and \textsc{javelin}) are then used on the mock continuum and line light curves to recover the input time lag. The results are shown in Figure \ref{Fig:mock_ccf} and given in Table \ref{Table:result_mock}.

 \begin{table}
 \caption{Time delay analysis results on mock light curve. Columns are as follows (1) method used, (2) input lag 70 days and (3) input lag 200 days between mock continuum and mock line light curve.}
	\begin{center}
 	\resizebox{1.0\linewidth}{!}{%
     \begin{tabular}{ l l l }\hline \hline 
     method                &  \multicolumn{2}{c}{lag} \\\hline
    input lag              & 70          &  200    \\
                           &  (days)                   &    (days)                          \\ 
      (1)                  &  (2)                      &    (3)                  \\ \hline
  ICCF                     & $ 70.0^{+1.9}_{-2.0}$     & Lag not calculated as $r_{max}<0.3$ \\
  DCF                      & $ 79.7^{+15.9}_{-20.2}$   & Lag not calculated as $r_{max}<0.3$ \\
  von Neumann              & $ 68.8^{+3.6}_{-1.1}$     & $217.5^{+16.9}_{-46.4}$\\
  Bartels                  & $ 69.3^{+5.5}_{-1.7}$     & $217.9^{+11.9}_{-43.9}$\\
  \textsc{javelin}         & $ 161.2^{+1.6}_{-89.6}$   & $200.4^{+53.4}_{-5.5}$\\        
     \hline 
        \end{tabular} } 
        \label{Table:result_mock}
        \end{center}
    \end{table}

\end{appendix}

\end{document}